\documentclass[twocolumn,10pt,showpacs,showkeys,preprintnumbers,aps,pre,letterpaper]{revtex4-1}
\usepackage{amsmath,amssymb,amsthm}
\usepackage{graphicx}
\usepackage{dcolumn}
\usepackage{bm}
\usepackage[english,num]{isodate}
\usepackage[pdftex]{hyperref}
\hypersetup{%
	pdftitle	= {Preisach models of hysteresis driven by Markovian input processes }, 
	pdfauthor	= {Sven Schubert},
	pdfproducer	= {Sven Schubert},
	pdfkeywords	= {Random processes, non-Gaussian processes, Hysteresis, Preisach model},
	pdfstartview= {FitB},
	pdffitwindow=true,
	bookmarksopen=false,
	pdfdisplaydoctitle=true
}

\newcommand*{\figref}[1]{Fig.\,\ref{#1}}
\newcommand*{\mean}[1]{\left< #1\right>}

\newcommand*{\sgn}{\text{sgn}}
\newcommand*{\tauinfty}{\tau\rightarrow\infty}
\newcommand*{\fnoise}{\mbox{$1\!/\!f$-noise }}
\newcommand*{\fanoise}{\mbox{$1\!/\!f^\alpha$-noise }}

\newlength{\figurewidth}
\newlength{\twofigurewidth}
\newlength{\figureheight}
\begin{document} 
\preprint{\isodate{\today}-svs-expdecay}

\title{Preisach models of hysteresis driven by Markovian input processes}

\author{Sven Schubert}
 \email{svs@physik.tu-chemnitz.de}
\author{G\"{u}nter Radons}%
\affiliation{%
Institute of Physics, Chemnitz University of Technology,
D-09107 Chemnitz, Germany
}%

\date{\today}

\begin{abstract}
We study the response of Preisach models of hysteresis
to stochastically fluctuating external fields.
%
We perform numerical simulations which indicate that
analytical expressions derived previously for the autocorrelation functions and
power spectral densities of the Preisach model with uncorrelated input,
hold asymptotically also if the external field shows exponentially decaying
correlations.
As a consequence, the mechanisms causing long-term memory and \fnoise in
Preisach models with uncorrelated inputs still apply in the presence of fast
decaying input correlations.
We collect additional evidence for the importance of the effective
Preisach density previously introduced even for Preisach models with correlated inputs.
%
Additionally, we present some new results for
the output of the Preisach model with
uncorrelated input using analytical methods. 
It is found, for instance, that 
in order to produce the same long-time tails in the output, the 
elementary hysteresis loops of large width need
to have a higher weight for the generic Preisach model
than for the symmetric Preisach model. 
Further, we find autocorrelation functions 
and power spectral densities 
to be monotonically decreasing
independently of the choice of input and Preisach density.

\end{abstract}

\pacs{02.50.Ey, 05.40.-a, 05.90.+m, 75.60.-d}
\keywords{random processes, \mbox{$1\!/\!f$}-noise, hysteresis, Preisach model}

\maketitle

\section{\label{sec:intro}Introduction} 
Hysteresis is 
a widespread phenomenon \cite{Bertotti2006a}
which is observed in nature and, moreover, the key feature of many technological applications.
It involves the development of a hysteresis memory, and multistability in the interrelations between external driving fields (input) and system response (output).
Prominent examples are the
magnetization of ferromagnetic materials in an external magnetic field \cite{Bertotti1998a},
or the adsorption-desorption hysteresis observed in porous media \cite{Everett1952}.
A phenomenological model which is successfully applied to many different systems with hysteresis is the Preisach model \cite{Mayergoyz1991,*Mayergoyz2003}.
Although, it was originally formulated for ferromagnetic materials \cite{Weiss1907,Preisach1935},
it was \citet{Everett1954} who realized its phenomenological character and the applicability to a wide
range of phenomena from different scientific fields.

Stochastically driven Preisach models were first investigated by 
\citeauthor{Mayergoyz1991b} \cite{Mayergoyz1991b,Mayergoyz1991c,Korman1994a,Mayergoyz1994a}.
To explain thermal relaxation processes and to describe the magnetic after-effect or creep phenomena, they suggested to study 
the relaxation of the average response 
of Preisach hysteresis models driven by discrete-time independent identically distributed (i.i.d.)\;continuous random processes of zero mean.
More recently, spectral properties of the response were investigated and a mechanism for the generation of long-term memory including \fnoise was revealed \cite{Radons2008a,Radons2008b,Radons2008c}.
The characterization of the memory configuration of Preisach models driven by diffusion and Ornstein-Uhlenbeck processes
is subject of research of \citet{Amann2011a}.
Aim of this paper is to
extend investigations for Preisach models driven by Ornstein-Uhlenbeck processes \cite{Dimian2004,Dimian2009a}
to non-Gaussian input processes in particular, with a focus on the long-term correlation decay of the system response.
%
In ref.\,\cite{Dimian2004}
a scheme was presented for the computation of the power spectral density of the output process
of Preisach models driven by Ornstein-Uhlenbeck processes.
%
For the same problem, power spectral densities were computed via Monte
Carlo simulations in ref.\,\cite{Dimian2009a}.
In the latter publication,
the hysteretic response is described for various hysteresis models including the
Preisach model.
It was found that the output spectra deviate significantly from the Lorentzian
shape of the spectrum of the input processes.
This is confirmed by our results presented here, but we go beyond
and show that, 
for input processes with exponentially decaying temporal correlations, 
output autocorrelation functions 
approach the asymptotic correlation decay already known for Preisach models driven by uncorrelated input processes \cite{Radons2008b}.
For the latter,
it was shown recently, using rigorous methods \cite{Radons2008a}, that
the development of a hysteresis memory is reflected in the possibility
of long-time tails in the autocorrelation functions of the system's response \cite{Radons2008c}.
These long-time tails represent a long-term memory where signal components with
arbitrary large periods of duration contribute significantly to the signal, which is reflected by the appearance of \mbox{$1\!/\!f^\alpha$-noise}.
The methods used to derive the rigorous results 
for models driven by uncorrelated processes are not applicable
to input processes showing temporal correlations,
as a consequence, we use simulation techniques.

This paper is organized as follows.
The Preisach model is described in Sect.\,\ref{sec:model}.
In Sect.\,\ref{sec:results}, we provide rigorous 
results on 
symmetric and
generic Preisach models driven by uncorrelated processes, some of it
recapitulated from ref.~\cite{Radons2008b,Radons2008a,Radons2008c}, 
followed by numerical results on correlated inputs in Sect.\,\ref{sec:results2}
and a brief summary in Sect.\,\ref{sec:sum}.

\section{\label{sec:model}Preisach model} 
\subsection{Definition}
\begin{figure}[!hb]
	\includegraphics{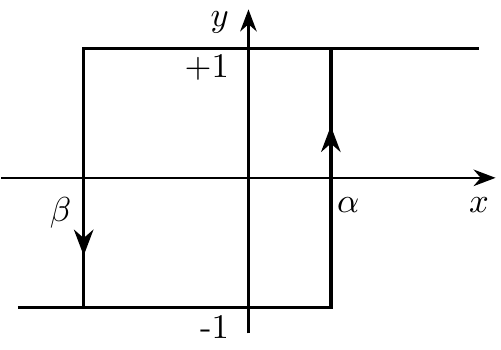} 
	\caption{%
		A Preisach unit $\hat{s}_{\alpha\beta}$ 
		with upper threshold $\alpha$ and lower threshold $\beta$. 
	}
	\label{fig:gamma}
\end{figure}

The Preisach model \cite{Mayergoyz1991,*Mayergoyz2003} generates
rate-independent hysteresis using independent domains.
The model is characterized by the Preisach operator $\hat\Gamma_\mu$, which acts on an input time series 
$(x_1,\ldots,x_t)$ and responds with an output time series 
$(y_1,\ldots,y_t)$.
The response of the Preisach operator $y_t=\hat\Gamma_\mu\{x_t\}$ is given by the weighted superposition of the response of Preisach units $\hat{s}_{\alpha\beta}$ acting on the external field, 
\begin{equation}\label{eq:Gamma}
\hat\Gamma_\mu\{x_t\} = \iint\limits_{\alpha\geq\beta}\! \text{d}\alpha \text{d}\beta
\,\mu(\alpha,\beta)\hat{s}_{\alpha\beta}\{x_t\} .
\end{equation}
A Preisach unit $\hat{s}_{\alpha\beta}$ is specified by its upper and lower threshold values $\alpha$ and $\beta$ determining an elementary rectangular hysteresis loop.
It yields the output $y_t=1$ if its input $x_t$ is 
larger than the upper threshold $\alpha$, and 
$y_t=-1$ if its input is less than the lower threshold $\beta$. 
In between, the Preisach unit is bistable, see Fig.~\ref{fig:gamma}.
Thus, the response of a Preisach unit can be written as
\begin{equation}\label{eq:gamma}
\hat{s}_{\alpha\beta}\{x_t\} = \left\{
\begin{array}{l}
+1 \text{~if~} \exists\, t_1\in[t_0,t]\!: \;
   x_{t_1}\geq \alpha ,\;x_\tau>\beta \;\forall\tau\in[t_1,t] \\
-1 \text{~if~} \exists\, t_1\in[t_0,t]\!: \;
   x_{t_1}\leq \beta ,\;x_\tau<\alpha \;\forall\tau\in[t_1,t] \\
s_{\alpha\beta}(t_0)\in\{-1,1\} \text{~if~} \beta < x_\tau < \alpha \;\forall \tau\in[t_0,t] \\
\end{array}
\right..
\end{equation}
$s_{\alpha\beta}(t_0)$ denotes the initial condition of the Preisach unit with threshold values $\alpha$ and $\beta$.
Each loop's individual weight is given by the Preisach density 
$\mu(\alpha,\beta)$.
In case one considers only symmetric Preisach units $\hat{s}_\alpha$ where $\beta=-\alpha$,
the Preisach density becomes
$\mu(\alpha,\beta)=\mu(\alpha)\delta(\alpha\!+\!\beta)$. Consequently, the
individual weights of the symmetric Preisach model are given by a function of one variable $\mu(\alpha)$.  

\subsection{Numerical simulations}

In the following, we summarize the most important aspects of the simulation and the used initial conditions.
The focus lies on the differences between the algorithm for the symmetric and
the generic case.

\subsubsection{The role of the initial configuration}
The latest output
and the system memory given by the history of the external field 
are not the only variables determining the system state. Its response, as well as its evolution, depends on the initial states $s_{\alpha\beta}(t_0)$ of Preisach units
as long as the input remains in the corresponding interval $(\beta,\alpha)$.
Even for a stationary random input process, the random output approaches a
stationary process only after the influence of the initial state has died out.
%
In case of symmetric input densities,
the use of an equilibrated initial state accelerates this behavior on average.
It is given by
\[
{s}_{\alpha\beta}(t_0)=\left\{
\begin{array}{ll}
+1 & \alpha\leq -\beta \\
-1 & \alpha > -\beta
\end{array}\right.
.
\]
In micromagnetics, this is called {the demagnetized state} \cite{Mayergoyz1991,*Mayergoyz2003}.

For symmetric Preisach models the corresponding equilibrated initial state is realized by 
$s_{\alpha}(t_0)= 0$. 
In this way, 
the responses of neighboring elementary hysteresis operators $\hat{s}_\alpha$ cancel each other out mimicking that they are in different states.

\subsubsection{\label{subsubsec:out_sim}The output computation}
Assuming the global absolute extreme value of the input time series in the time interval $[t_0,t]$ 
is a minimum $m_0=\min\limits_{\tau\in[t_0,t]}\{x_\tau\}$,
the computation of the output $y_t$ could be carried out by summing up
integrals over triangular regions 
\begin{equation}\label{eq:sum_out}
y_t = 2\sum\limits_{i=1}^{N}\big[
F(M_i,m_{i-1})-F(M_i,m_{i})\big]-F(M_0,m_0)
\end{equation}
where $M_0=m_0$ and
$
F(\alpha,\beta)=\int\limits_\beta^\alpha \text{d}\alpha '\int\limits_\beta^{\alpha '} \text{d}\beta '\mu(\alpha ',\beta ').
$
The values stored by the model are a reduced sequence of return points consisting of decreasing local input maxima $M_i$ 
and increasing local minima $m_i$. This sequence is called 
an alternating series of dominant extreme values.
Their number of entries $N$ varies with time. 
One finds an expression analogous to Eq.~(\ref{eq:sum_out}) in case
the global absolute extreme value of the input time series is a maximum $M_0$,
\begin{equation*}
y_t = F(M_0,m_0) - 2\sum\limits_{i=1}^{N}\big[
F(M_{i-1},m_i)-F(M_i,m_{i})\big].
\end{equation*}
For more details on the system's memory and the computation of the system response see ref.~\cite{Mayergoyz1991,*Mayergoyz2003}.

The algorithm for symmetric Preisach models where $\mu(\alpha,\beta)=\mu(\alpha)\delta(\alpha\!+\!\beta)$
is slightly different.
The fact that this Preisach density is concentrated on the line $\beta=-\alpha$
simplifies the computation of the output.
The values stored by the model divide this line of Preisach units into line
segments with units in the up state, $\hat{s}_\alpha(x_t)=+1$, and segments with units in the down state, $\hat{s}_\alpha(x_t)=-1$.
For each time step, the output is performed 
using the expressions of the cumulative Preisach distribution function
$F(\alpha)=\int\limits_0^\alpha d\alpha '\mu(\alpha ')$.
One sums up the contributions from all segments with Preisach units in the up state given by $F(M_{i+1})-F(m_{i})$ and from segments with Preisach units in the down state given by $F(m_i)-F(M_i)$. The difference of both is the system's output. Here, $M_i>0$ and $m_i<0$ are the different return points of the input  memorized by the system at the time $t$.

\section{\label{sec:results}Hysteretic systems driven by uncorrelated input processes} 
\subsection{\label{subsec:symmetric}The autocorrelation
function of symmetric Preisach models}
In this section, 
we give more compact expressions for the computation of autocorrelation
functions and power spectral densities which 
deviate from the presentation in ref.\,\cite{Radons2008b}.
Further, 
we will document, among other things, that our numerical methods reproduce the
analytical results \cite{Radons2008a,Radons2008b,Radons2008c} for the asymptotic
decay of the output autocorrelation function of Preisach models driven by
uncorrelated inputs.  

First we treat symmetric Preisach models driven by uncorrelated symmetric input processes $\{X_t\}$,
where $p_X(x)=p_X(-x)$ and $F_X(x)=\int_{-\infty}^x\text{d}x'p_X(x')$ denote the
input probability density and the cumulative distribution function, respectively.
We recapitulate and extend some of the results on the output processes $\{Y_t\}$ from ref. \cite{Radons2008b}.

There exists an effective Preisach density given by 
\begin{equation}\label{eq:def_mueff}
	\tilde\mu(u)=\frac{\mu[\alpha(u)]}{2p_X[\alpha(u)]}
	\text{~with~}
	u = 2[1-F_X(\alpha)]. 
\end{equation}
The support of $\tilde{\mu}$ is the interval $[0,1]$.
Thus the effective Preisach density $\tilde\mu$ captures in one function the
combined effect of the Preisach density $\mu$ and the input density $p_X$.
In terms
of this density the stationary output autocorrelation functions 
$
C_Y(\tau) = \lim\limits_{t\rightarrow\infty}
\left(\mean{Y_tY_{t+\tau}}-\mean{Y_t}\mean{Y_{t+\tau}}\right)
$
can be expressed as
\begin{equation}
C_Y(\tau) =
\int\limits_{0}^1 \!\text{d}u\,\tilde\mu(u)
\int\limits_{0}^1 \!\text{d}{u'}\,\tilde\mu(u')
\frac{\min(u,u')}{\max(u,u')}
(1-u')^{|\tau|} .~~~
\label{eq:corz_mueff}
\end{equation}
This expression 
follows from Eqs.\,(17) -- (19) presented in ref.~\cite{Radons2008a}.
It 
follows from results for the cross-correlation function of the output of two
Preisach units. The state of two Preisach units is governed by a 4-state
Markovian process determined by a transition matrix with entries governed by the
cumulative distribution function of the input process.
Eq.\,(\ref{eq:corz_mueff}) shows that the output autocorrelation function is
given by a superposition of exponential correlation decays,
\begin{equation}
\label{eq:corz_exp}
C_Y(\tau) =
\int\limits_{0}^\infty \!\text{d}\lambda\, g(1-e^{-\lambda})e^{-\lambda|\tau|}
\end{equation}
where $g(u)  = 
(1-u)\tilde\mu(u)\int_0^1\text{d}u'\,\tilde\mu(u')
\frac{\min(u,u')}{\max(u,u')}$.
As a consequence of Eq.\,(\ref{eq:corz_mueff}), all systems with identical effective Preisach densities show the same autocorrelation function. 
Moreover, we can show that systems which have an identical effective Preisach
density yield realizations of the same stochastic output process, i.\,e. any
two combinations of input and Preisach density resulting in the same effective
Preisach density yield two output processes for which not only correlations
but all compound probability densities coincide, see App.\,\ref{subsec:eff_den}.
Further, we can show easily that the autocorrelation function is monotonically
decreasing for $\tau \geq 0$; taking the derivative of
Eq.\,(\ref{eq:corz_exp}) with respect to $\tau$ yields
\begin{equation*}
\frac{\text{d}}{\text{d}\tau}C_Y(\tau) = - \sgn (\tau)
\int\limits_{0}^\infty \!\text{d}\lambda\,
g(1-e^{-\lambda})\lambda\, e^{-\lambda|\tau|}.
\end{equation*}
Thus, for positive Preisach densities and, therefore, positive effective %
densities $\tilde\mu(u)$, one sees easily that $\frac{\text{d}}{\text{d}\tau}C_Y(\tau)\leq 0$ $\forall\,\tau\geq 0$. 
In addition, 
the curvature of $C_Y(\tau)$ is positive for finite values of $\tau$,
which can be concluded from the second derivative.
Further, we will give an expression for the power spectral density which
follows by a Fourier transform (Wiener-Khinchin theorem)
which reads
\[
S_Y(\omega) = 2\,\Re\left\{
\sum\limits_{\tau=0}^\infty C_Y(\tau) e^{\text{i}\omega\tau}
\right\}-C_Y(0) 
\]
due to the symmetry of $C_Y(\tau)$.
Using Eq.\,(\ref{eq:corz_mueff}) and the geometric series, 
it follows that the power spectral density can be computed by
\begin{equation}
S_Y(\omega) =
\int\limits_{0}^1 \!\text{d}u\,\tilde\mu(u)
\int\limits_{0}^1 \!\text{d}{u'}\,\tilde\mu(u')
\frac{\min(u,u')}{\max(u,u')}
\frac{1-q^2}{1-2q\cos\omega+q^2}
\label{eq:psdz_mueff1D}
\end{equation}
where $q=1-u'$.
This expression is a compressed version of Eq.\,(9) presented in ref.~\cite{Radons2008b} which
follows from Eq.\,(22) in ref.~\cite{Radons2008a}.
Its derivative with respect to $\omega$ yields
\begin{eqnarray*}
\frac{\text{d}}{\text{d}\omega}S_Y(\omega) &=& -
\int\limits_{0}^1 \!\text{d}u\,\tilde\mu(u)
\int\limits_{0}^1 \!\text{d}{u'}\,\tilde\mu(u')
\frac{\min(u,u')}{\max(u,u')} \\
& & \times
\frac{2q(1-q^2)}{(1-2q\cos\omega+q^2)^2}\sin\omega ,
\end{eqnarray*}
which results in $\frac{\text{d}}{\text{d}\omega}S_Y(\omega)\leq 0$ where $\omega\in [0,\pi]$;
the power spectral densities of symmetric Preisach models driven by 
a sequence of i.i.d.\;random variables (uncorrelated input process) decay monotonically
regardless of Preisach and input density.
The monotonicity of the spectral density and its independence of the shape of the Preisach density was already expected for symmetric Preisach models driven by Ornstein-Uhlenbeck input processes \cite{Dimian2004}.
Further, depending on the effective density,
there is a possibility of $S_Y(\omega)$ to have a point of inflection 
where the second derivative changes sign.

For input and Preisach densities which belong to the same class of functions, for instance
for asymptotically algebraically decaying input and Preisach densities
\begin{eqnarray}
\label{eq:p_algden}
p_X(x) & = & \frac{\nu}{2}(1+|x|)^{-(\nu+1)},\;\nu >0.\\
\label{eq:mu_algden}
\mu(\alpha) & = & \nu '(1+\alpha)^{-(\nu '+1)},\;\nu '>0,
\end{eqnarray}
one obtains from Eq.\,(\ref{eq:def_mueff}) the following effective Preisach density
\begin{equation}\label{eq:mueff_ex}
\tilde\mu(u)=\gamma_1 u^{\gamma_1-1} \text{~with~} u\in [0,1] 
\text{~and~} \gamma_1=\frac{\nu '}{\nu}>0.
\end{equation}
This effective Preisach density is also obtained for a pair of Pareto
densities, a pair of exponential densities, and a pair of densities with
algebraic behavior on limited support $p_X(x) =
\nu/2(1-|x|)^{\nu-1},\;\nu>0,\,-1<x<1$ and $\mu(\alpha) = \nu
'(1-|\alpha|)^{\nu'-1},\;\nu '>0,\,0\leq\alpha<1$. Furthermore, the effective
Preisach density which belongs to a pair of Gaussian densities can be
approximated by Eq.\,(\ref{eq:mueff_ex}) with additional logarithmic
corrections.
The evaluation of the integrals in Eq.\,(\ref{eq:corz_mueff})
yields the explicit result
\begin{equation}\label{eq:corz_example}
C_Y(\tau)=\left\{
\begin{array}{ll}
	\frac{H_{\tau+2}-1/2}{2+3\tau+\tau^2} & \gamma_1=1\\
	\gamma_1^2\Gamma(\tau\!+\!1)\left[
	\frac{\gamma_1\Gamma(\gamma_1\!-\!1)}{\Gamma(\tau\!+\!\gamma_1\!+\!2)}
	+\frac{2\,\Gamma(2\gamma_1)}{(1-\gamma_1^2)\Gamma(\tau\!+\!2\gamma_1\!+\!1)}
	\right] & \text{else}
\end{array}
\right. .
\end{equation}
$H_n$ is the $n$-th harmonic number and $\Gamma(z)$ is the $\Gamma$-function.
The autocorrelation function (\ref{eq:corz_example}) shows the following asymptotic behavior
\begin{equation}\label{eq:corz_asymp_example}
C_Y(\tau)\sim
\left\{
	\begin{array}{ll}
		\frac{2\gamma_1^2\Gamma(2\gamma_1)}{(1-\gamma_1^2)}\tau^{-2\gamma_1} & 0<{\gamma_1}<1 \\
		\tau^{-2}{\ln\tau} & \gamma_1=1 \\
		\gamma_1^3\Gamma(\gamma_1\!-\!1)\tau^{-(\gamma_1+1)} & \gamma_1>1
	\end{array}
\right. \;(\tauinfty)
\end{equation}
which is in accordance with Eq.\,(23) from ref.~\cite{Radons2008b}.
Additionally, we see in Eq.\,(\ref{eq:corz_example}) the short term behavior,
which is used to test the numerical tools used later to produce the results for
correlated input scenarios.
%
The output autocorrelation functions show an asymptotically algebraic decay 
$C_Y(\tau)\sim\tau^{-\eta_\delta}$ with decay exponent
\[
\eta_\delta=\left\{
	\begin{array}{ll}
		2\gamma_1  & 0<{\gamma_1}<1 \\
		\gamma_1+1 & \gamma_1>1
	\end{array}
\right. .
\]
Furthermore, one observes \fanoise if $0<\gamma_1<1/2$, i.e.\;%
the power spectral density $S_Y(\omega)$
diverges as $\omega\!\rightarrow\! 0$
and the autocorrelation function is not absolutely summable, $\sum\limits_\tau|C_Y(\tau)|\rightarrow\infty$,
such that the system has developed long-term memory.
This scenario is similar to van der Ziel's explantion of \fnoise
in semiconductors \cite{Ziel1950} using a superposition of exponentially
decaying correlations with different decay rates, cp. Eq.\,(\ref{eq:corz_exp}).
Concluding, the Preisach model of hysteresis is able to transform uncorrelated
input into output with long-time tails in its autocorrelation function \cite{Radons2008c}.
In case the autocorrelation function of the output shows any algebraic decay,
the $n$-th derivative of $S_Y(\omega)$ with respect to $\omega$ 
is diverging \cite{Radons2008a}.
Consequently,
$S_Y(\omega)$ is nonanalytic at $\omega\rightarrow 0$
where the degree of nonanalyticity $n$ is given by
$
n=\lceil\min\{2\gamma_1-1,\gamma_1\}\rceil.
$
$\lceil\gamma_1\rceil$ denotes the smallest integer greater than or equal to $\gamma_1$.
The presented results hold asymptotically 
for all effective Preisach densities with 
\[
\tilde\mu(u)\sim u^{\gamma_1-1} (u\rightarrow 0)
\]
since the small $u$-behavior caused by elementary hysteresis loops of large width
determines 
the long-term correlation decay, see Eq.\,(\ref{eq:corz_mueff}).

In case the Preisach density belongs to a ``class of broader functions'' than the input density, the output becomes, apart from logarithmic corrections, \fnoise where $S_Y(\omega)\sim 1/\omega$ ($\omega\rightarrow 0$).
In case the input density belongs to the ``class of broader functions'' the
output correlation decays exponentially.

%

\subsection{\label{subsec:nonsymmetric}The autocorrelation
function of generic Preisach models}
%
To present some results on generic Preisach models driven by 
uncorrelated processes in a condensed fashion,
we follow the steps from the previous section.

There is an effective Preisach density \cite{Radons2008c}
\begin{equation}\label{eq:def_mueff2D}
	\tilde\mu(u,v)=\frac{\mu[\alpha(u),\beta(v)]}{p_X[\alpha(u)]p_X[\beta(v)]}
\end{equation}
with $u = 1-F_X(\alpha)$ and $v = F_X(\beta)$.
Systems with the same effective Preisach density show the same stationary autocorrelation function, which
follows from
\begin{eqnarray}\nonumber
C_Y(\tau) &=& 4
\int\limits_{0}^1 \!\text{d}u\!\int\limits_{0}^{1-u} \!\text{d}v\,\tilde\mu(u,v)
\int\limits_{0}^1 \!\text{d}{u'}\!\int\limits_{0}^{1-u'} \!\text{d}{v'}\,\tilde\mu(u',v') \\
& & \times
\frac{\min(u,u')\min(v,v')}{(u+v)(u'+v')}
(1-u'-v')^{|\tau |}
,\label{eq:corz_mueff2D}
\end{eqnarray}
cp. Eqs.\,(17) - (19) presented in ref.~\cite{Radons2008a}.
The corresponding power spectral density
of the system response follows
\begin{eqnarray}\nonumber
S_Y(\omega) &=& 4
\int\limits_{0}^1 \!\text{d}u\!\int\limits_{0}^{1-u} \!\text{d}v\,\tilde\mu(u,v)
\int\limits_{0}^1 \!\text{d}{u'}\!\int\limits_{0}^{1-u'} \!\text{d}{v'}\,\tilde\mu(u',v') \\
& & \times
\frac{\min(u,u')\min(v,v')}{(u+v)(u'+v')}
\frac{1-q^2}{1-2q\cos\omega+q^2}
\label{eq:psdz_mueff2D}
\end{eqnarray}
where $q=1-u'-v'$.
From the derivatives of Eqs.\,(\ref{eq:corz_mueff2D}) and (\ref{eq:psdz_mueff2D}) follows that
autocorrelation functions ($\forall\tau\leq 0$) and power spectral densities
($\omega\in[0,\pi]$) of the response of generic Preisach models to uncorrelated
processes decay also monotonically.

For input densities 
\begin{equation}\label{eq:pdf2}
p_X(x) = \nu (1-|x|)^{\nu -1},\;\nu >0,
\end{equation}
where $x\in(-1,1)$
and a constant Preisach density $\mu(\alpha,\beta)=1/2$ for $-1\leq\beta\leq\alpha\leq 1$,
one obtains an effective Preisach density as follows
\[
\tilde\mu(u,v)=2\gamma_2^2 [4\min(u,1\!-\!u)\min(v,1\!-\!v)]^{\gamma_2-1} 
\]
where $\gamma_2={1}/{\nu}>0$.
%
It follows from Eq.\,(\ref{eq:corz_mueff2D}) that
the asymptotic behavior close to the origin of ordinates $(u,v)\rightarrow (0,0)$ gives the significant contribution to the long-term correlation decay.
For the chosen example one can write 
\begin{equation}\label{eq:mueff2D_ex}
\tilde\mu(u,v)\propto u^{\gamma_2-1}v^{\gamma_2-1},
\end{equation}
which gives an effective Preisach density analogous to the example shown above
for the symmetric Preisach model, Eq.\,(\ref{eq:mueff_ex}).
The contributions of the symmetric Preisach units are given by
\begin{equation}\label{eq:mueff2D_uu}
\tilde\mu(u,u)\propto u^{\gamma_2-2}.
\end{equation}
App.\,\ref{subsec:eff_den} reveals that
it is sufficient to look only on the effective Preisach density.
Systems which show the same effective Preisach density $\tilde\mu(u,v)$ yield
realizations of the same output process $\{Y_t\}$. Consequently, the particular
form of input and Preisach density does not matter as long as they result in
the same effective Preisach density.

Using polar coordinates $u=r\sin\varphi$ and $v=r\cos\varphi$
and performing the angle integration, one derives an expression $\tilde\rho(r)$ similar to the effective density of the symmetric problem, Eq.\,(\ref{eq:mueff_ex})
\begin{eqnarray}
\tilde\rho(r) &:=&
\int\limits_0^{{\pi}/{2}}\!\text{d}\varphi\,r\tilde\mu(r\sin\varphi,r\cos\varphi) 
\nonumber \\ \label{eq:mueff_corr1D}
\tilde\rho(r) &\propto & 
r^{2\gamma_2-1}\quad (r\rightarrow 0).
\end{eqnarray}
Hence, neglecting logarithmic corrections, we expect the asymptotic correlation
decay of the generic Preisach model driven by an uncorrelated input process to
follow from the corresponding symmetric problem with $\gamma_1=2\gamma_2$.
%
\begin{figure}[ht]
	\centering
\parbox[c]{1.2\twofigurewidth}{
\includegraphics[width=1.2\twofigurewidth]{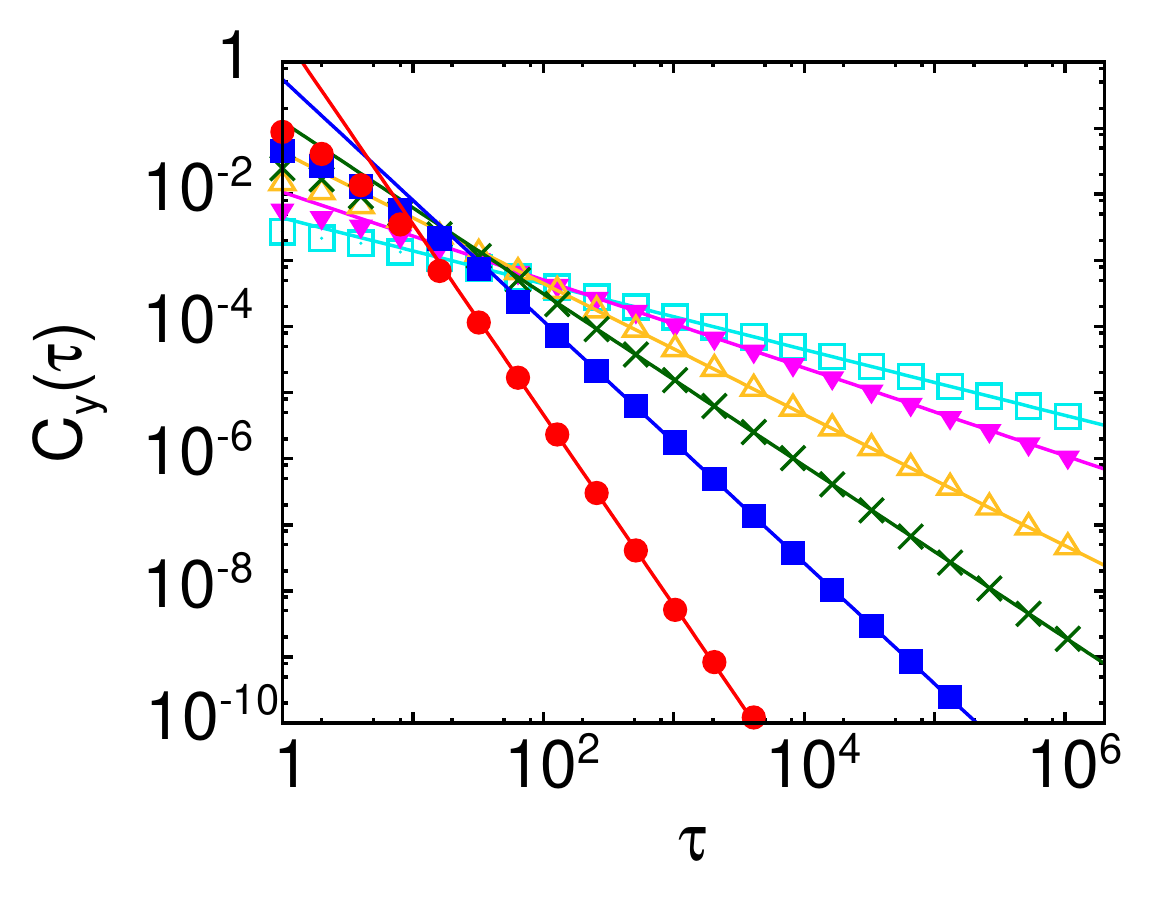} 
}\hspace{4mm}
\parbox[c]{0.5\twofigurewidth}{
\begin{tabular}{c|c|c}
\hline\hline 
$\gamma_2$ & $\eta_\text{fit}$ & $\eta_\delta$ \\
\hline 
1/8 & 0.498 & 1/2 \\
1/6 & 0.666 & 2/3 \\
1/4 & 0.992 & 1 \\
1/3 & 1.301 & 4/3 \\
1/2 & 1.87 & 2 \\
1 & 2.83 & 3 \\
\hline\hline 
\end{tabular}
\vspace{20pt}
}
	\caption{%
The autocorrelation function of the response of a generic Preisach model to
uncorrelated input processes 
is fitted with a power law.
The different $\gamma_2$-values chosen are listed in the table
in the same order (from top to bottom) as in the diagram at large $\tau$-values.
The corresponding decay exponents $\eta_\text{fit}$ used to fit the data (solid
lines) and the exponents $\eta_\delta$ predicted by the approximation, see
Eq.\,(\ref{eq:corz_asymp_ex2D}), are given as well.
	}
\label{fig:Cor_P2D_uni_pl}
\end{figure}
The solution to which is known, see Eq.\,(\ref{eq:corz_asymp_example}). 
Thus, we expect an algebraic correlation decay $C_Y(\tau)\sim
\tau^{-\eta_\delta}$, see Fig.\,\ref{fig:Cor_P2D_uni_pl}, where 
\begin{equation}\label{eq:corz_asymp_ex2D}
\eta_\delta =
\left\{
	\begin{array}{ll}
{-4\gamma_2} & 0<{\gamma_2}<1/2 \\
{-(2\gamma_2+1)} & \gamma_2>1/2
	\end{array}
\right. \;(\tauinfty)
.
\end{equation}
The points in \figref{fig:Cor_P2D_uni_pl} belong to the autocorrelation
functions of generic Preisach models driven by uncorrelated processes where the
behavior of the effective Preisach density is determined
by the $\gamma_2$-value, see Eq.\,(\ref{eq:mueff2D_ex}).
They follow from the numerical evaluation of Eq.\,(\ref{eq:corz_mueff2D}).
The $\gamma_2$-values chosen are listed in the table shown next to
the diagram. 
Further, the decay exponents $\eta_\text{fit}$ used to fit the data by a power
law and the exponents $\eta_\delta$ predicted by the approximation,
Eq.\,(\ref{eq:corz_asymp_ex2D}), are given.
The stronger the correlation decay the more marked is the tendency to
underestimate the decay exponent, which explains the deviations between fit and
prediction in Fig.\,\ref{fig:Cor_P2D_uni_pl}.  
For the case $\gamma_2 =1$, the asymptotic correlation decay was calculated 
analytically in ref.~\cite{Radons2008a}. 
The result, logarithmic corrections to an algebraic decay with the decay exponent $\eta_\delta=3$, is matched by the approximation given in Eq.\,(\ref{eq:corz_asymp_ex2D}).

\begin{figure}[t]
	\centering
	\textbf{(a)}\hspace{-6mm}%
\includegraphics[width=\figurewidth]{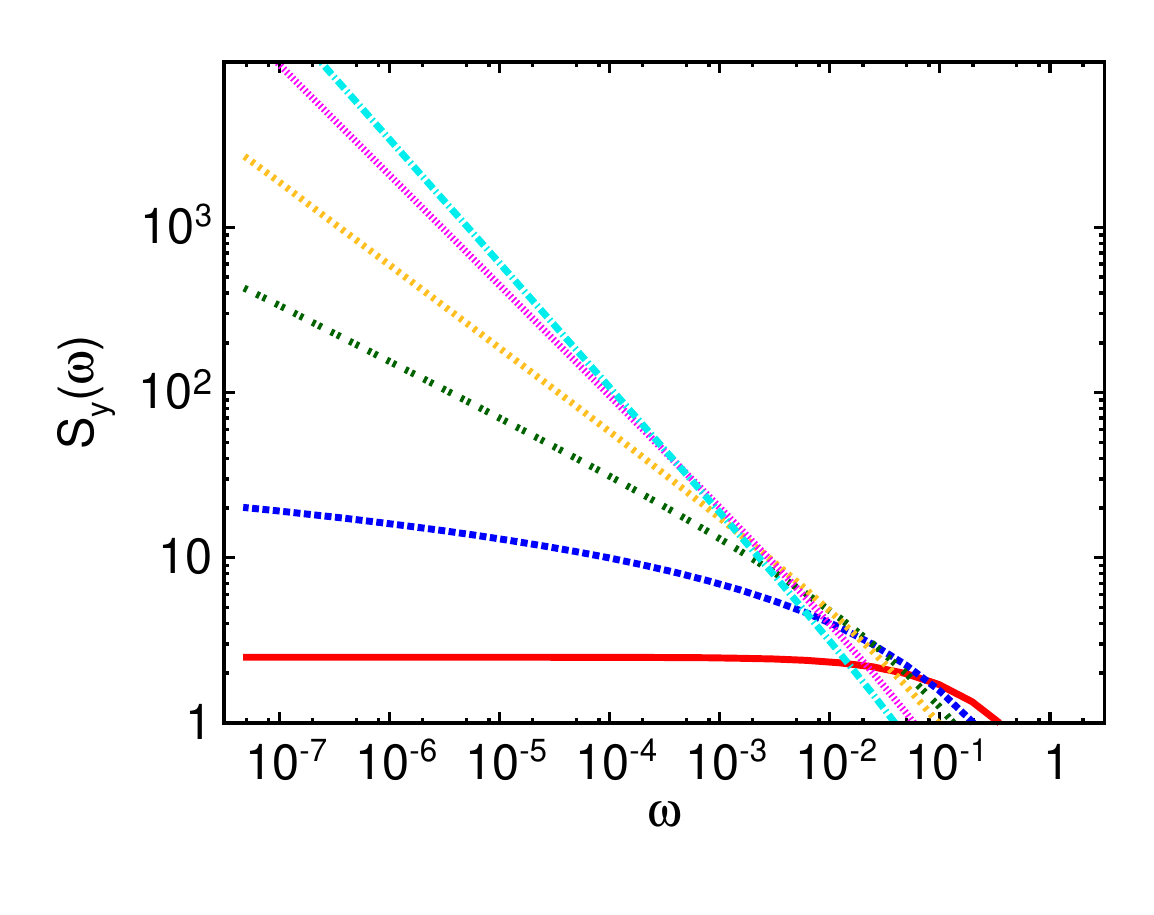} \\
	\textbf{(b)}\hspace{-6mm}%
\includegraphics[width=\figurewidth]{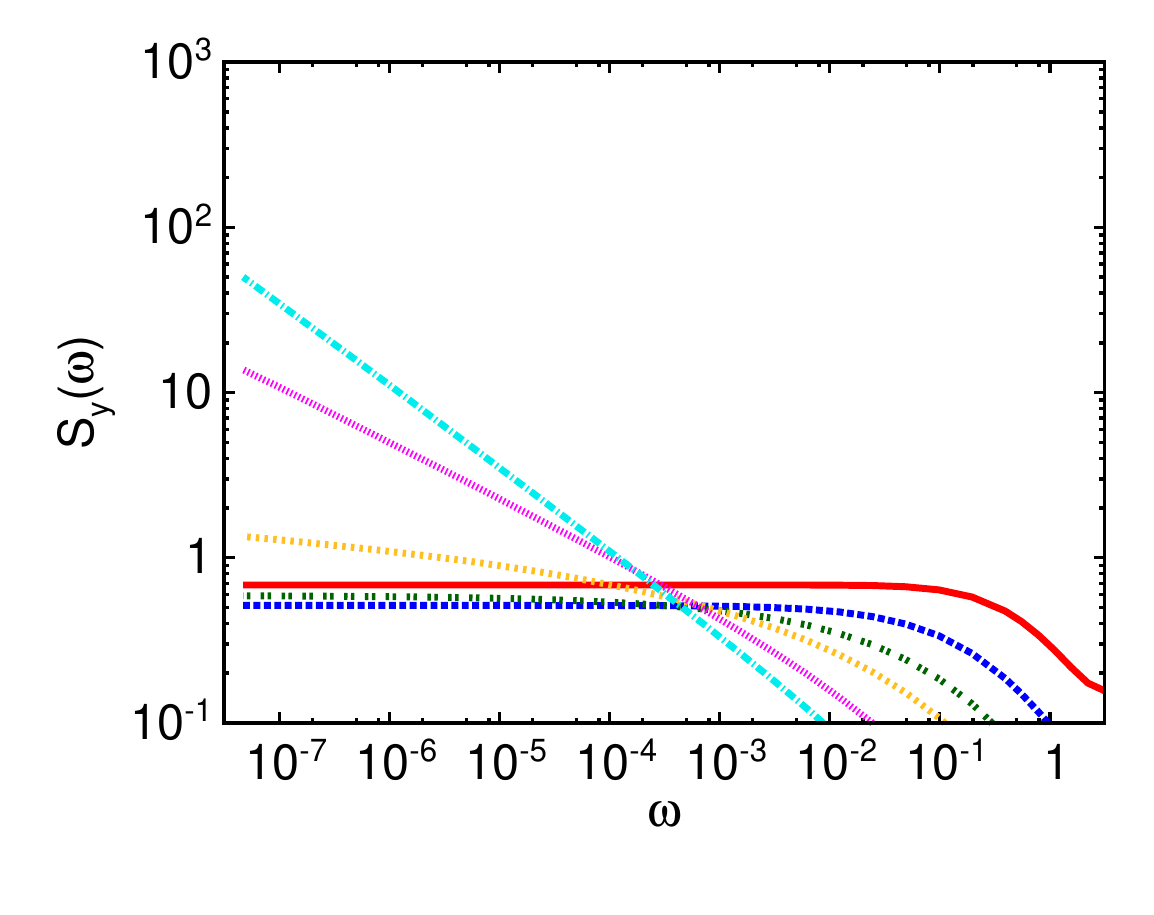}
	\caption{%
The analytic results of the power spectral densities are shown for symmetric
\textbf{(a)} and generic \textbf{(b)} Preisach models driven by 
uncorrelated processes. 
The parameters determining the effective Preisach densities
take the values
$\gamma_1 = \gamma_2 =1,1/2,1/3,1/4,1/6$ and $1/8$.
Smaller values $\gamma$ correspond to a stronger increase of the power spectral
density as $\omega$ approaches 0.
	}
\label{fig:PSD_cp}
\end{figure}

There are contributions of elementary loops of large width
besides the contribution of symmetric Preisach units,
Eq.\,(\ref{eq:mueff2D_uu}).
Consequently, the behavior of the generic model is determined by an average of
these contributions, which mimicks here a symmetric Preisach model whose
elementary loops of large width have less weight than the symmetric contributions of the
generic model.
Also from Eq.\,(\ref{eq:corz_asymp_ex2D}), we expect \fanoise 
and therefore long-term memory 
in the system response for $\gamma_2<1/4$. 
Fig.\,\ref{fig:PSD_cp}\,a) and b) provide a comparison of the power spectral
densities of the symmetric case and the generic case. 
The effective Preisach densities are given by Eqs.\,(\ref{eq:mueff_ex}) and
(\ref{eq:mueff2D_ex}), respectively.
The power spectral densities are computed using
Eqs.\,(\ref{eq:psdz_mueff1D}) and (\ref{eq:psdz_mueff2D}), respectively.
%
%

\subsection{\label{subsec:out_pdf}The output density of symmetric
Preisach models}
We will present results for the output distribution of a symmetric Preisach
model with uncorrelated input. The system is determined by the effective
Preisach density given by Eq.\,(\ref{eq:mueff_ex}).
At first, we will provide rigorous results for the variance. Secondly, we will
compare numerical results for the output density $p_Y(y)$ with an empirical
formula.
The output density is solely determined by the effective Preisach density since 
all combinations of input and Preisach density resulting in the same effective
Preisach density yield output processes for which even compound probability
densities coincide.

The output variance follows from 
\begin{equation}\label{eq:var_example}
\text{Var}(Y_t)=C_Y(\tau=0)=\frac{\gamma_1}{\gamma_1+1}.
\end{equation}
We take a look at two limits: a broad input density and a broad Preisach
density.
In case of a broad input density 
where $\gamma_1\rightarrow 0$,
the variance goes towards zero. Consequently, the output density approaches a
Dirac $\delta$-function $p_Y(y)\rightarrow \delta(y)$.
The second limit 
is given by $\gamma_1\rightarrow \infty$. 
From Eq.\,(\ref{eq:var_example}) follows
that the variance approaches $1$. Since the output is bounded on the interval
$[-1,1]$, the output density approaches two delta peaks
$p_Y(y)\rightarrow\frac{1}{2}\big[\delta(y\!-\!1)+\delta(y\!+\!1)\big]$.
The process behaves like a uncorrelated spin variable.
This behavior is obvious since $\gamma_1\rightarrow \infty$ corresponds to 
$\mu(\alpha)\rightarrow\delta(\alpha)$.
Thus, the Preisach model returns the sign of the uncorrelated symmetric input process.
Between both limiting cases, the variance monotonically increases with
increasing $\gamma_1$.
The output density $p_Y(y)$ has to become broader as the input density $p_X(x)$
becomes broader.
Such behavior corresponds to a correlation
decay with decreasing decay exponent since broader elementary loops are
less pronounced.

This behavior is reflected by simulations. 
The following results are for fixed $\nu=5/2$ and different values $\nu '$,
see Eqs.\,(\ref{eq:p_algden}) and (\ref{eq:mu_algden}).

\begin{figure}[t]
	\centering
	\textbf{(a)}\hspace{-6mm}%
\includegraphics[width=1.0\figurewidth]{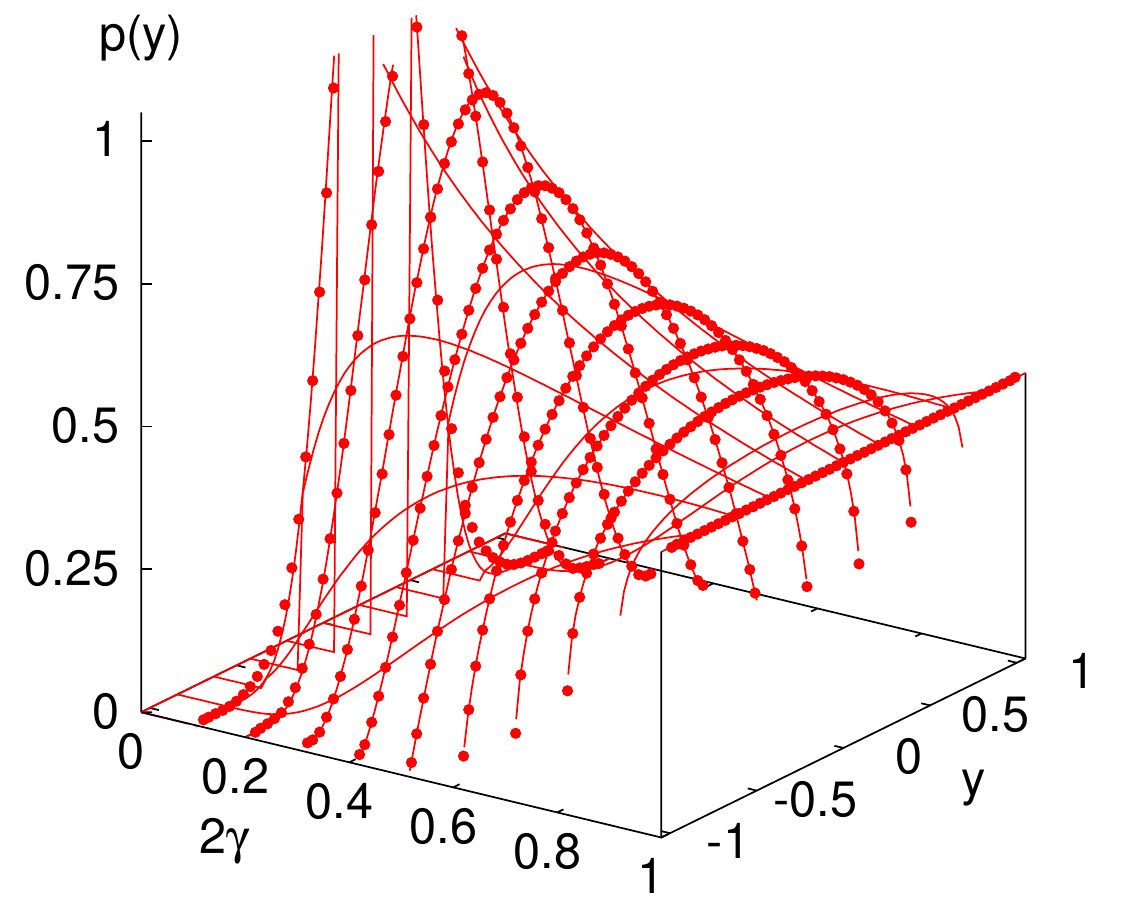}\\
\textbf{(b)}\hspace{-6mm}%
\includegraphics[width=1.0\figurewidth]{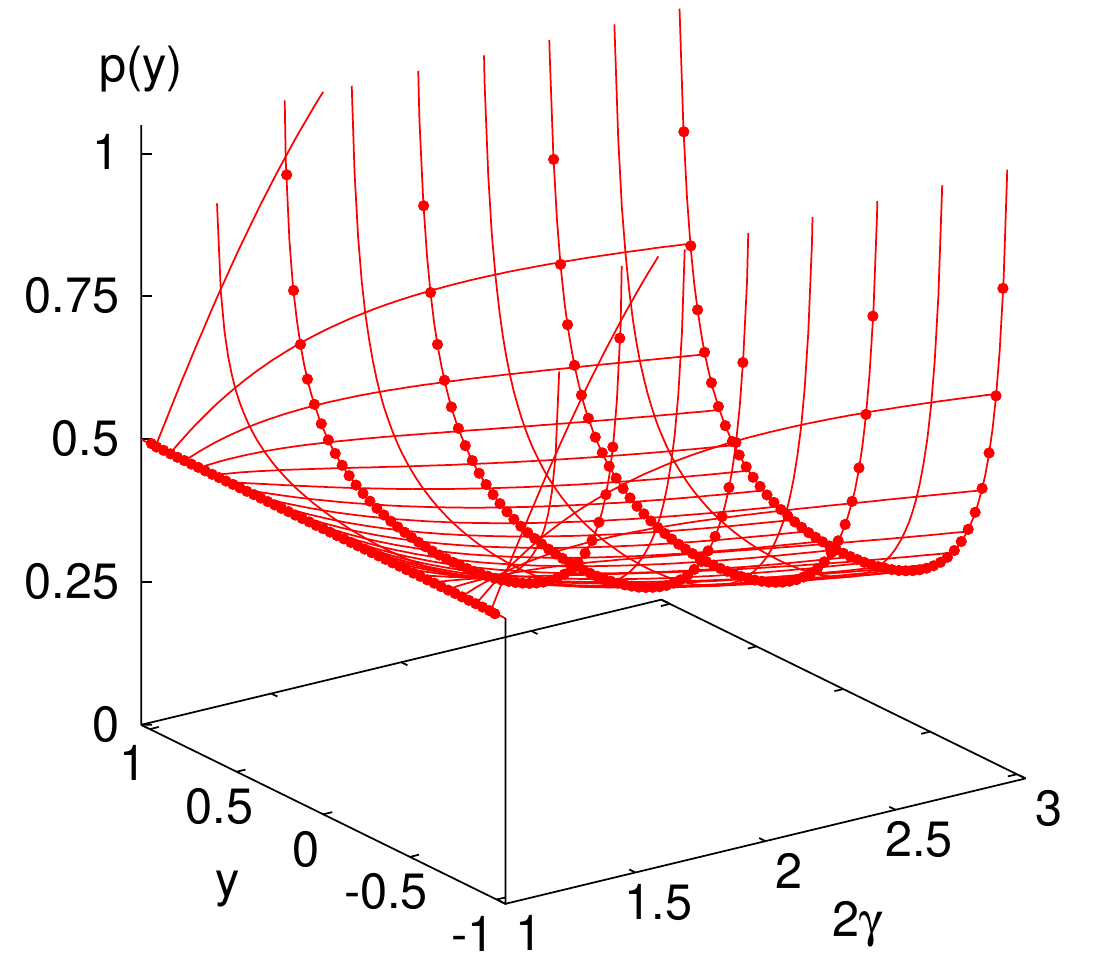}
	\caption{%
		The solid lines belong to the output density according to
		Eq.\,(\ref{eq:PDFy}). The points are estimates based on a simulation.
		As the Preisach density becomes broader, the parameter $\gamma_1$ decreases
		and 
		the corresponding output density 
		becomes more narrow. 
		\textbf{(a)} For $\gamma_1<1/2$ the output density shows a peak and vanishes
		as $y$ approaches the saturation values $y=\pm 1$.
		\textbf{(b)} For $\gamma_1>1/2$ the output density becomes bimodal and
		diverges
		at the borders.
	}\label{fig:PDFy_3D}
\end{figure}

\begin{figure}[t]
	\centering
\begin{tabular}{rr}
	\textbf{(a)}\hspace{16mm} & \textbf{(c)}\hspace{16mm} \\
	\includegraphics[width=\twofigurewidth]{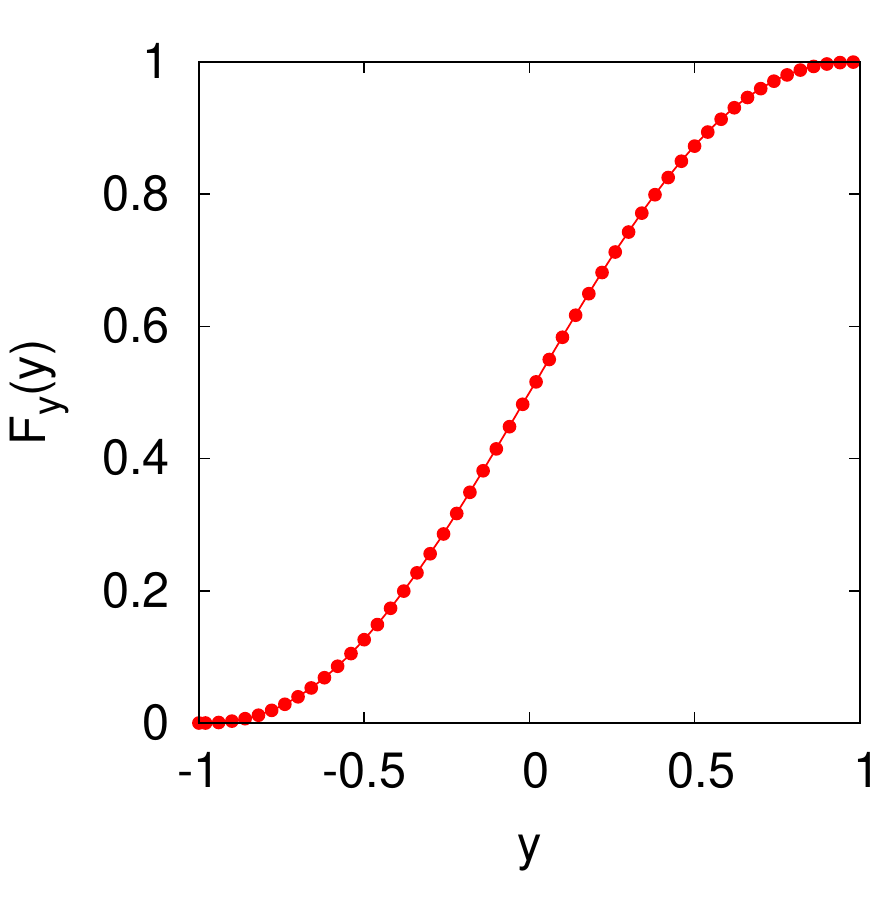}
	&
	\includegraphics[width=\twofigurewidth]{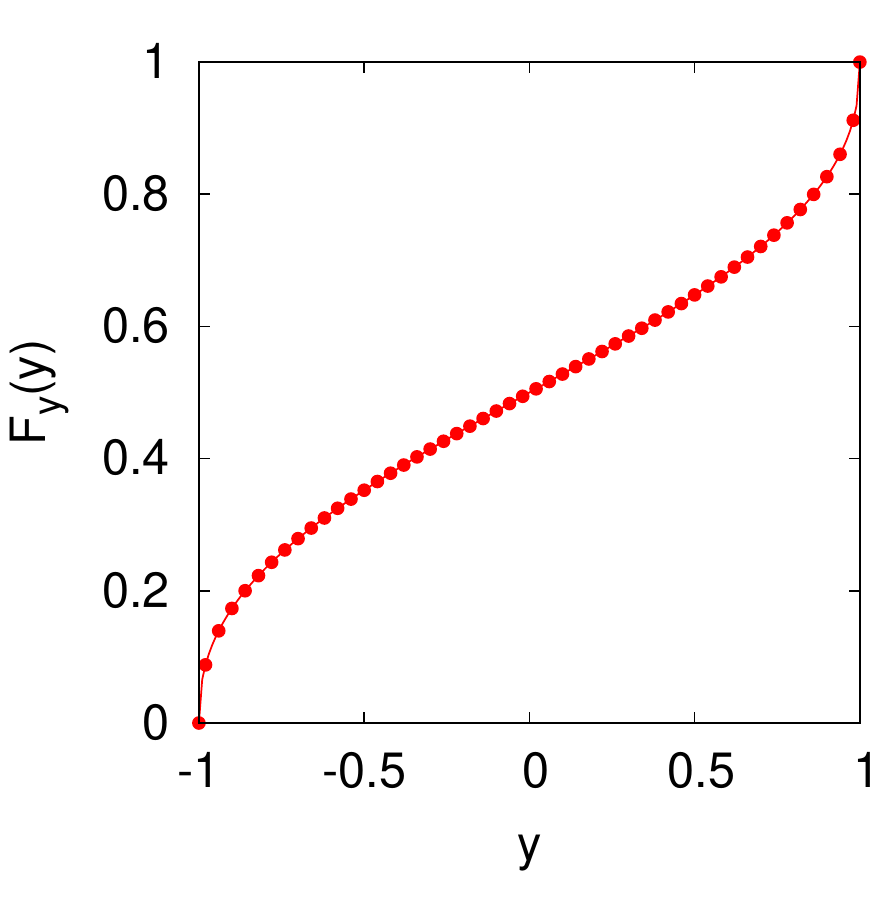}
	\\
	\textbf{(b)}\hspace{16mm} & \textbf{(d)}\hspace{16mm} \\%
	\includegraphics[width=\twofigurewidth]{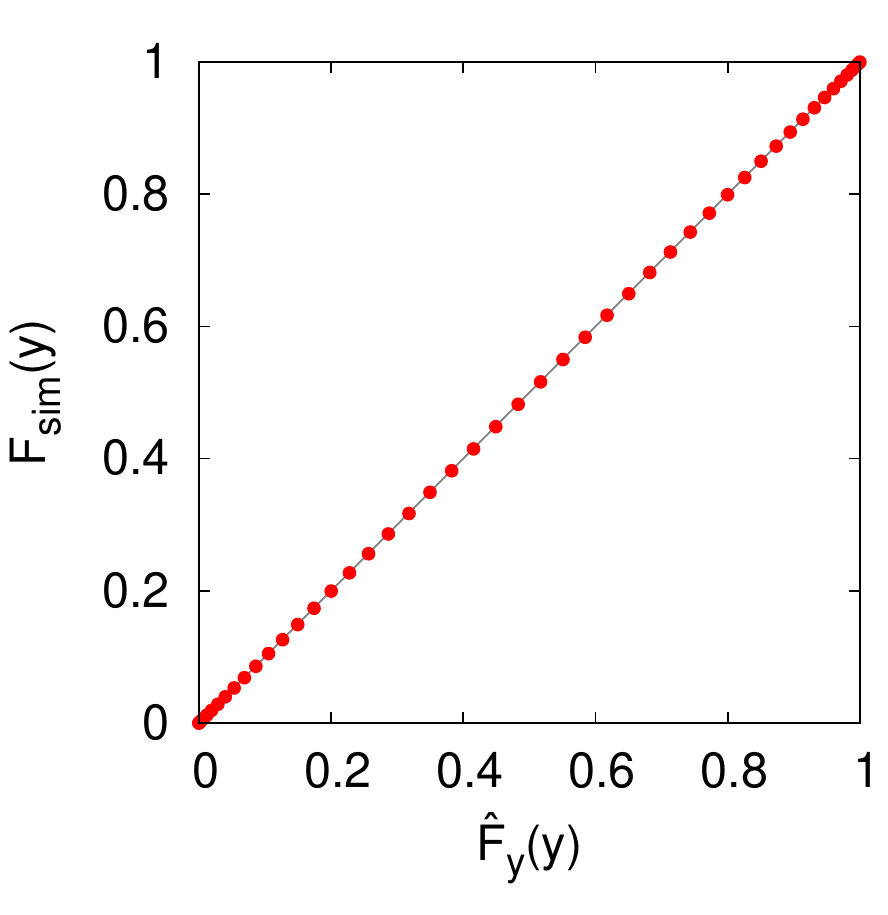} 
 	&
	\includegraphics[width=\twofigurewidth]{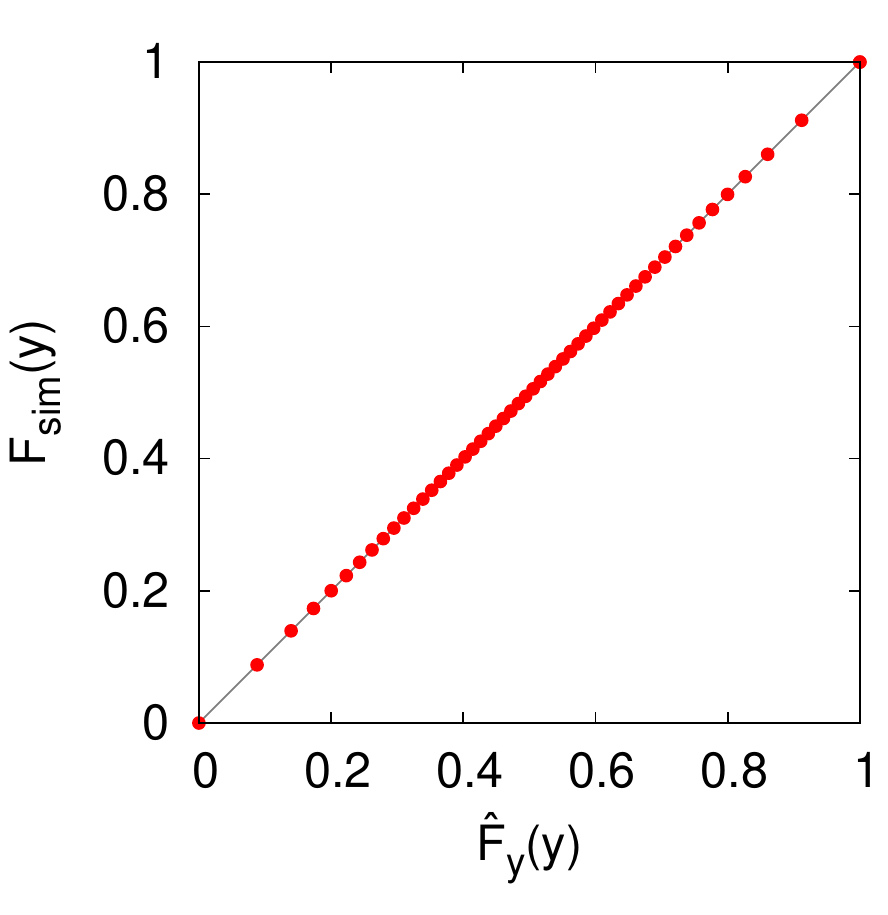} 
\end{tabular}
	\caption{%
		The figures \textbf{(a)} and \textbf{(b)} belong to $\gamma_1=1/5$;
		the figures \textbf{(c)} and \textbf{(d)} belong to $\gamma_1=6/5$.
		They present 
		the cumulative distributions, \textbf{(a)} and \textbf{(c)}, 
		and P-P plots, \textbf{(b)} and \textbf{(d)}, of 
		symmetric Preisach models with uncorrelated input.
		The models are determined by Eq.\,(\ref{eq:mueff_ex}).
	    The solid lines base on the empirical formula, Eq.\,(\ref{eq:PDFy}), the
	    points belong to the simulation.
	  	}\label{fig:PPplot}
\end{figure}

The output density can be approached by a shifted Beta distribution with the
following empirical formula 
\begin{equation}\label{eq:PDFy}
	\hat p(y)=
	\frac{\Gamma(a\!+\!1/2)}{\Gamma(a)\,
	\sqrt{\pi}}\big(1-y^2\big)^{a-1},
	\quad a=\frac{1}{2\gamma_1} .
\end{equation}
The corresponding cumulative distribution function is given mainly by an
incomplete Beta function.
Unfortunately, we are currently not able to prove Eq.\,(\ref{eq:PDFy}), but an
extremely 
good agreement between simulations and the empirical formula is
documented in \figref{fig:PDFy_3D}.
First of all, the corresponding variance matches Eq.\,(\ref{eq:var_example}). 
Secondly, we are going to substantiate the above statement by P-P plots which
plot the cumulative distribution functions estimated from the data against the
cumulative distribution functions according to Eq.\,(\ref{eq:PDFy}).
Numerical data are compared with results for the cumulative distribution
function according to Eq.\,(\ref{eq:PDFy}) for $\gamma=1/5$ and $\gamma=6/5$,
Fig.\,\ref{fig:PPplot}.

\section{\label{sec:results2}Hysteretic systems driven by correlated input processes} 
To obtain results for stochastically driven hysteresis from numerical
experiments, we first need to generate stochastic input processes
$(X_1,X_2,\ldots)$ with given probability density $p_X(x)$
and stationary autocorrelation function 
$C_X(\tau)=\lim_{t\rightarrow\infty}\big(\mean{X_t\,X_{t+\tau}}-\mean{X_t}\mean{X_{t+\tau}}\big)$ 
with given asymptotic correlation decay.
This is done in two steps. Firstly, the generation of an AR(1)-process provides 
the exponential correlation decay. Secondly, 
a monotonic transformation is performed to obtain the cumulative distribution function 
of the target processes and, consequently, its probability density.
The algorithm used is presented in detail in App.\,\ref{sec:input}.

\subsection{\label{subsec:symmetric2}The autocorrelation function of
symmetric Preisach models}
\begin{figure}[t]
	\centering
\includegraphics[width=\figurewidth]{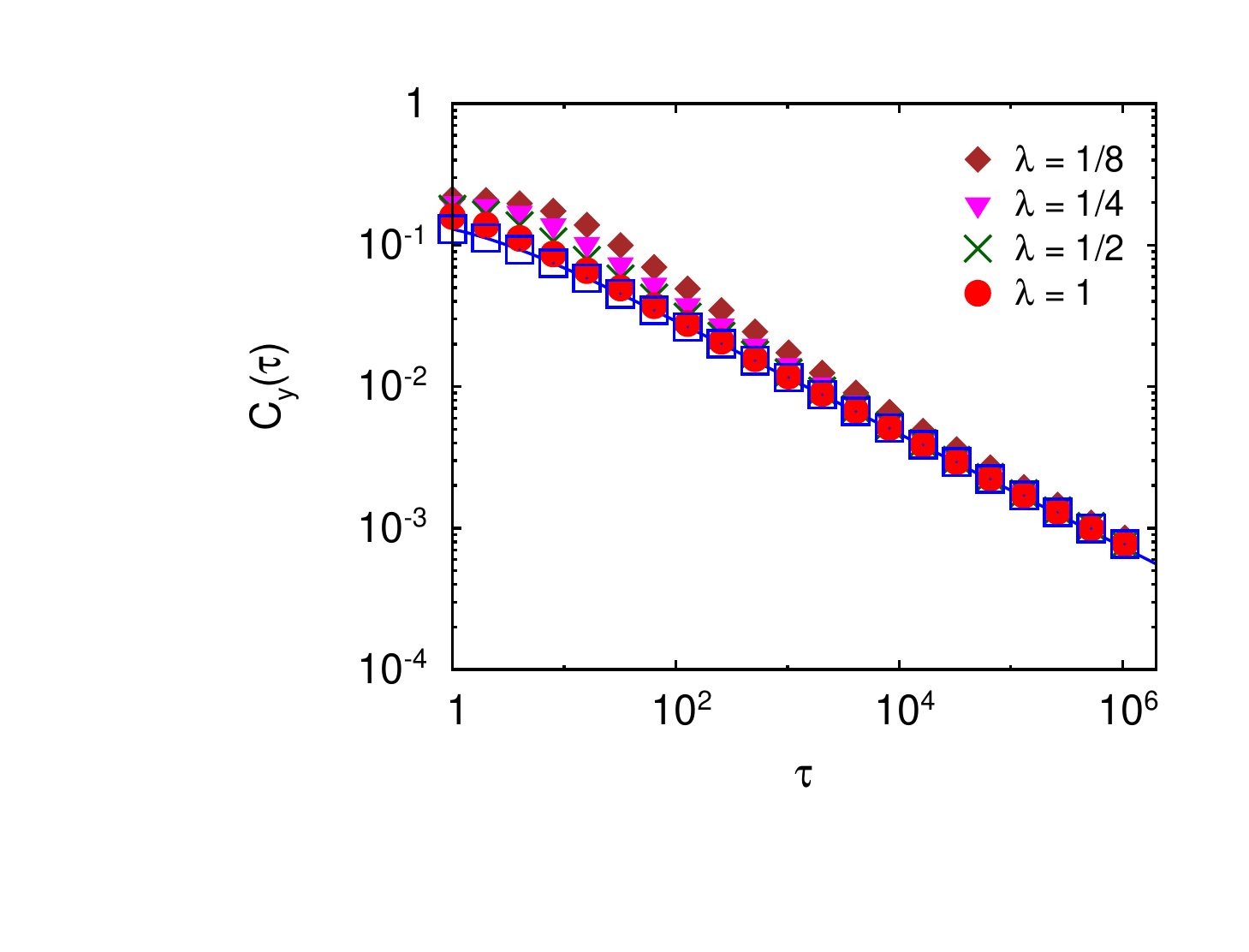}
	\caption{%
		The autocorrelation function of the output
		is shown for symmetric Preisach models ($\gamma_1=1/5$)
		driven by Markovian input processes with exponentially 
		decaying autocorrelation functions where 
		$\lambda=1,1/2,1/4$ and $1/8$. 
		The output correlation decay  
		yields the same asymptotic behavior as the corresponding situation 
		with uncorrelated driving, simulation (blue boxes) and rigorous results 
		(blue line), see Eq.\,(\ref{eq:corz_example}).
	}
\label{fig:cor_P1Deq_alg2_tf01_AR}
\end{figure}

\begin{figure}[ht]
	\centering
\includegraphics[width=\figurewidth]{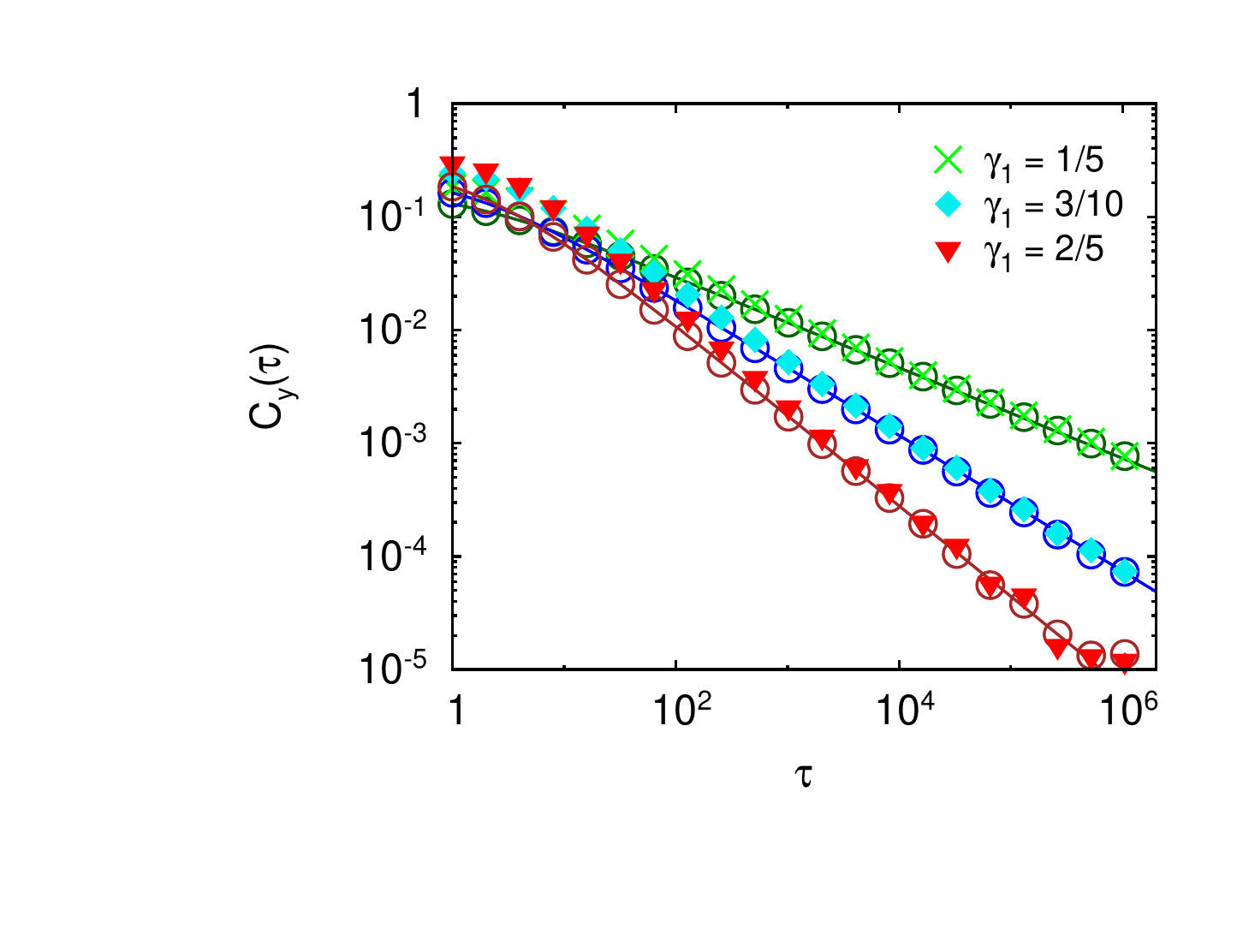}
	\caption{%
		The autocorrelation function of the output
		is shown for different symmetric Preisach models 
		($\gamma_1=2/10,3/10$ and $4/10$) which are
		driven by Markovian input processes with exponentially 
		decaying autocorrelation functions 
		where $\lambda=1/2$.
		The autocorrelation functions approach the corresponding 
		correlation decay for uncorrelated driving (circles and line) asymptotically.
	}
\label{fig:cor_P1Deq_alg_tf01_AR02}
\end{figure}

In the following, we consider symmetric Preisach models with exponentially
decaying input correlations
\begin{equation}\label{eq:cor_exp}
C_X(\tau) \sim e^{-\lambda\,\tau}\quad (\tau\rightarrow\infty)
\end{equation}
and an input density given by $p_X(x)$. 
We compute the output autocorrelation function
$C_y(\tau)=\overline{y_ty_{t+\tau}}-\overline{y_t}^2$ 
using time and ensemble averages.
An estimate of the autocorrelation function at large times only from ensemble
average would require large ensembles and would be of much larger numerical
effort. The latter can be reduced significantly by the additional time
average.
Note that the data shown in
Fig.\,\ref{fig:Cor_P2D_uni_pl},
\ref{fig:PDFy_3D} - \ref{fig:cor_P1Deq_alg_tf01_AR02}, 
\ref{fig:cor_P2Deq_uni_plH_AR} - \ref{fig:cor_tf01_AR}
are obtained from averaging over 1024 samples of length $2^{24}$.
The good agreement of analytical expressions (solid lines) and simulation data
(open symbols) for Preisach models with uncorrelated inputs supports the assumption that the procedure
chosen is suitable. 

At first, we chose the example of a driven symmetric Preisach model 
introduced previously in Sect.\,\ref{subsec:symmetric} 
where
Preisach and input densities decay asymptotically algebraically, Eqs.\,(\ref{eq:p_algden}) and (\ref{eq:mu_algden}).
For uncorrelated inputs, the behavior of the system is determined by 
the quotient of the decay exponents $\gamma_1=\frac{\nu '}{\nu}$.
The input time series computed show a finite correlation decay rate $\lambda$,
see Eq.\,(\ref{eq:cor_exp}).

Two scenarios are investigated. Firstly, the parameters determining the input
and Preisach densities are fixed, $\nu=5/2$ and $\nu'=1/2$ ($\gamma_1=1/5$), 
and the input correlation decay rate $\lambda$ takes different values.
The output autocorrelation functions which follow from the simulations using
$\lambda=1,1/2,1/4$ and $1/8$ are shown in \figref{fig:cor_P1Deq_alg2_tf01_AR}. 
Additionally, \figref{fig:cor_P1Deq_alg2_tf01_AR} contains
the numerical data and the rigorous result, Eq.\,(\ref{eq:corz_mueff}), of the corresponding scenario where the symmetric Preisach model is driven by an uncorrelated process.
The finite decay rates $\lambda$ in the input autocorrelation functions cause a
slightly higher level of the output autocorrelation functions at short time
scales, which tends to decay faster until it approaches the correlation decay
already known for Preisach models driven by uncorrelated input processes.
As a consequence, the autocorrelation decay is determined by the effective
Preisach density's parameter $\gamma_1$ even in the presence of exponentially
decaying input correlations. That implies the influence of input and Preisach
density is reduced to the influence of the effective Preisach density, see
Eq.\,(\ref{eq:def_mueff}), for the asymptotic output correlation decay.

Secondly, we keep the asymptotic input correlation decay rate fixed, $\lambda=1/2$, and look at systems with different Preisach densities
such that the parameter $\nu'$ and, as a consequence, $\gamma_1$ is varied.
\figref{fig:cor_P1Deq_alg_tf01_AR02} shows that the long-term correlation decay is not affected by the finite decay rate $\lambda$ of the input signal. Asymptotically, the same behavior is observed as in the case of uncorrelated driving, hence $C_y(\tau)\sim\tau^{-\eta}$ ($\tauinfty$) where $\eta \approx \eta_\delta$.

\begin{figure}[t]
	\centering
\includegraphics[width=\figurewidth]{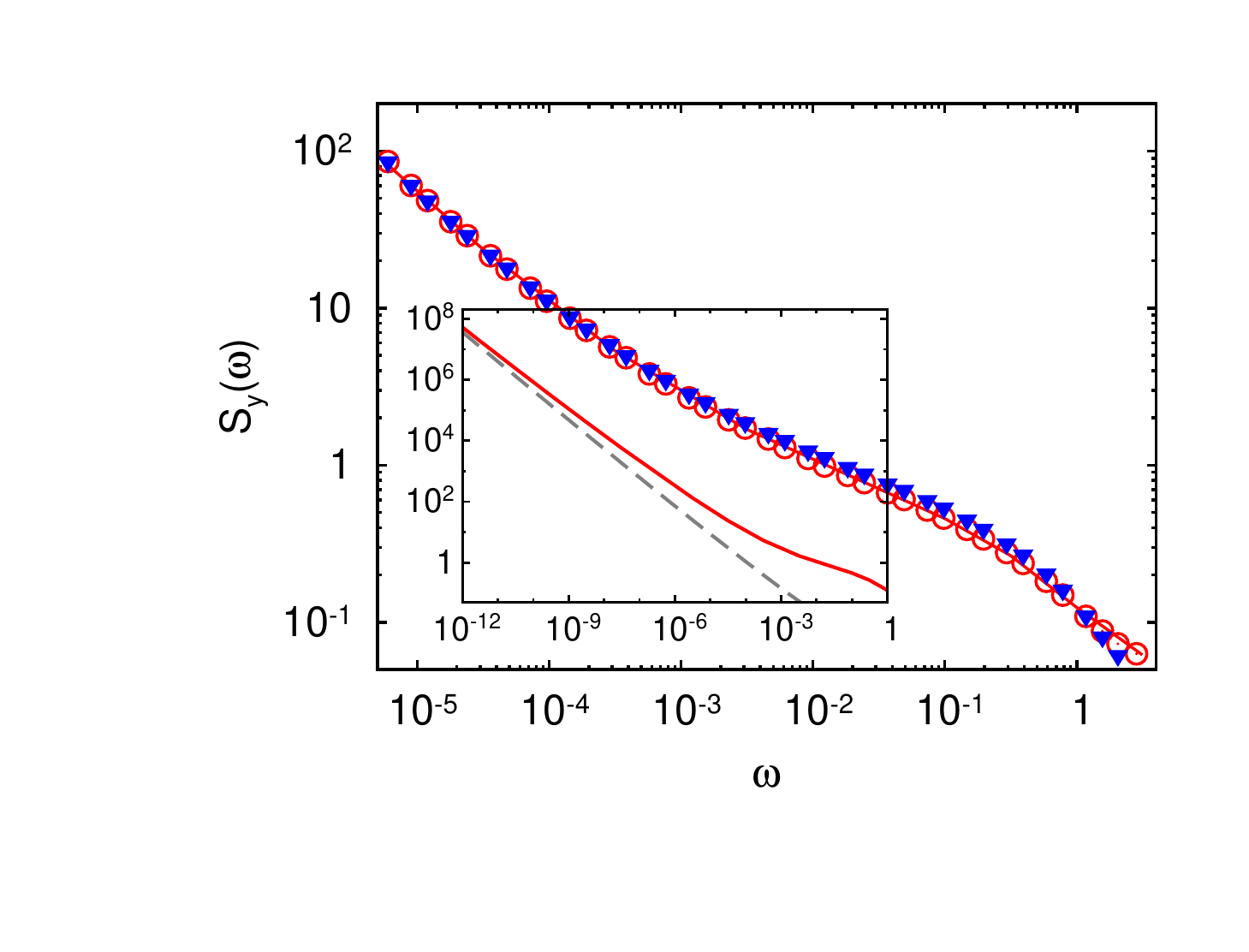}
	\caption{%
The power spectral density is obtained by a simulation of 
a symmetric Preisach model with a broad Preisach density 
driven by an uncorrelated process (red circles) and an exponentially
correlated input processes (blue triangles), both with a narrow probability density.
The prediction of \fnoise is verified by the inset 
providing a comparison with 
$S(\omega)\propto -\frac{1}{\omega\log\omega}$ (gray dashed line).
	}
\label{fig:PSD_P1Deq_alg7_GN}
\end{figure}

\begin{figure}[t]
	\centering
\includegraphics[width=\figurewidth]{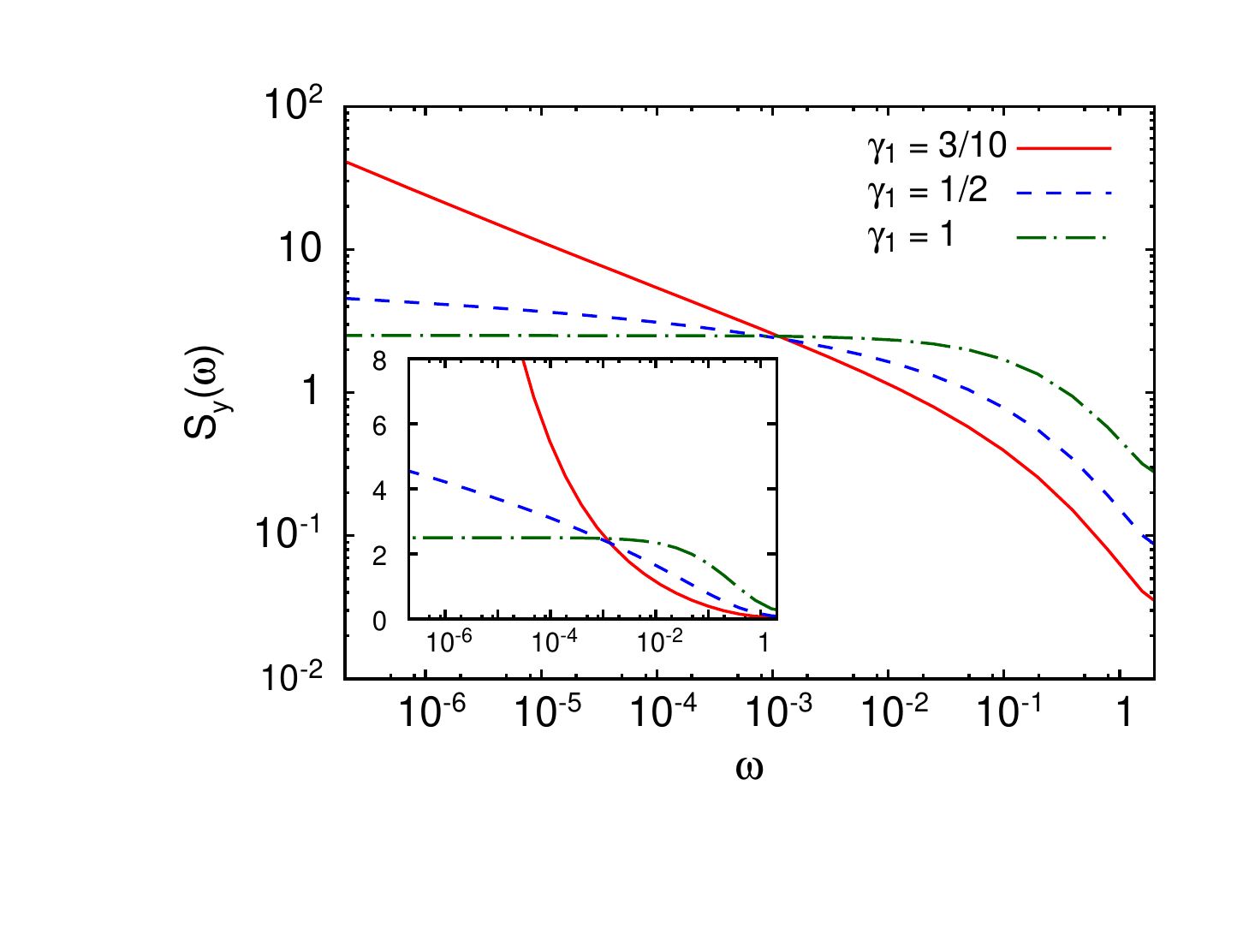}
	\caption{%
The figure shows the power spectral densities of symmetric Preisach models
with Gaussian input and Preisach density.
The power spectral density shows an almost Lorentzian shape, a logarithmic, and
a power law divergence depending on the model parameters.
The divergence is illustrated by the log-log plot and the semi-logarithmic
plot in inset. 
	}
\label{fig:PSD_P1D_GaussGauss}
\end{figure}

The fact that broad Preisach and narrow input densities
yield \mbox{$1\!/\!f$-noise}, where $S_y(\omega)\sim 1/\omega$ ($\omega\rightarrow 0$),
is displayed in \figref{fig:PSD_P1Deq_alg7_GN}.
The power spectral density is computed by an ensemble average
$S_y(\omega)=\frac{1}{N}\overline{|{\cal F}\{y_t\}(\omega)|^2}$
where $N$ denotes the length of the output time series and ${\cal F}\{\cdot\}$ the Fourier transform.
For the latter we used the Hann window function.
The figure shows the power spectral density obtained from a simulation of 
a symmetric Preisach model with an asymptotically algebraically decaying
Preisach density, Eq.\,(\ref{eq:mu_algden}), driven by two different Gaussian
random processes.
The power spectral density of the response to an uncorrelated process 
is also asymptotically 
approached using an exponentially correlated Markovian input process 
with a finite correlation decay rate $\lambda=1/2$. 
The red solid line follows from Eq.\,(\ref{eq:psdz_mueff1D}) and predicts
\fnoise asymptotically, apart from logarithmic corrections.
The power spectral density thus obtained is nonanalytic at $\omega=0$
and the degree of nonanalyticity is zero. 
%
%
%
The autoregressive model of order 1, which we used here, is the discrete-time
analogue of the Ornstein-Uhlenbeck process.
\citet{Dimian2004} plotted spectral densities $S_y(\omega)$
of symmetric Preisach models with uniform and Gaussian Preisach density driven by Ornstein-Uhlenbeck processes.
Their figures suggest an almost Lorentzian shaped spectral density which
corresponds to an exponential decay of the output autocorrelation function and
show no signs of long-term memory.
In the case of a (narrow) uniform Preisach density on limited support,
the Gaussian input density represents the much broader distribution.
Consequently, the behavior observed is well understood by the conclusions drawn
in the previous section, which state
that the output correlation decays exponentially if the input density belongs to
a ``class of broader functions'' than the Preisach density.
However, the behavior of the output autocorrelation function changes drastically
if we have a Gaussian input density and, additionally, a Gaussian Preisach
densitiy with variance $\sigma^2$ and variance $\sigma'^2$, respectively. 
Neglecting the influence of the exponentially fast decay of the input
autocorrelation function, the system's behavior is determined by the effective
Preisach density with $\gamma_1=\sigma^2/\sigma'^2$.
In detail, the effective Preisach density is
given by
\begin{equation*}
\tilde\mu(u) = \gamma_1\, e^{(1-\gamma_1)\,\text{erfc}^{-1}(u)^2}.
\end{equation*}
Here, $\text{erfc}^{-1}(\cdot)$ denotes the inverse of the complementary error
function.
The power spectral density then behaves for $\omega\rightarrow 0$ as:
\begin{equation*}
S_y(\omega)\sim\left\{
\begin{array}{rl}
\omega^{-(1-2\gamma_1)} & 0<\gamma_1 < 1/2 \\
-\log(\omega) & \gamma_1 = 1/2 \\
\text{const.} & \gamma_1 > 1/2
\end{array}
\right. \,(\omega\rightarrow 0).
\end{equation*}
Besides a finite value $S_Y(\omega=0)$ \cite{Dimian2004}, we would also expect a
logarithmic and a power law divergence depending on the parameters chosen, see
\figref{fig:PSD_P1D_GaussGauss}.
Since
our results for the correlation decay of Preisach models with
uncorrelated inputs apply to the output correlation decay of Preisach models
with exponentially correlated inputs asymptotically ($\tau\rightarrow\infty$),
see \figref{fig:cor_P1Deq_alg_tf01_AR02},
the power spectral densities of Preisach models
with exponentially correlated inputs behave asymptotically
($\omega\rightarrow 0$) like the power spectral densities of Preisach models
with uncorrelated inputs.
Thus, our results suggest \fanoise or a certain degree of nonanalyticity of the
ouput power spectral density, which is a novelty to the field and was not found
before, neighter analytically \cite{Dimian2004} nor by simulation
\cite{Dimian2009a}.
In distinction from the Ornstein-Uhlenbeck processes considered in ref.\,\cite{Dimian2004},
the processes discussed here showed no bias, i.e.\;$\mean{X_t}=0$.
The role of a bias to the noise mean value is not extensively discussed so far. 
However, the asymptotic results presented here apply to most input densities 
since the tail of the density remains under an additional shift.

\subsection{\label{subsec:nonsymmetric2}The autocorrelation
function of generic Preisach models}
Again, we take a look at the generic Preisach model with a constant Preisach
density $\mu(\alpha,\beta)=1/2$, $-1\leq\beta\leq\alpha\leq 1$; the model is
driven by a stochastic process with the probability density given by
Eq.\,(\ref{eq:pdf2}) and a finite correlation decay rate $\lambda$. 
The parameter determining the effective Preisach density, the quotient of the 
decay exponents, becomes now $\gamma_2=1/{\nu}$, see Eq.\,(\ref{eq:mueff2D_ex}). 

\begin{figure}[t]
	\centering
\includegraphics[width=\figurewidth]{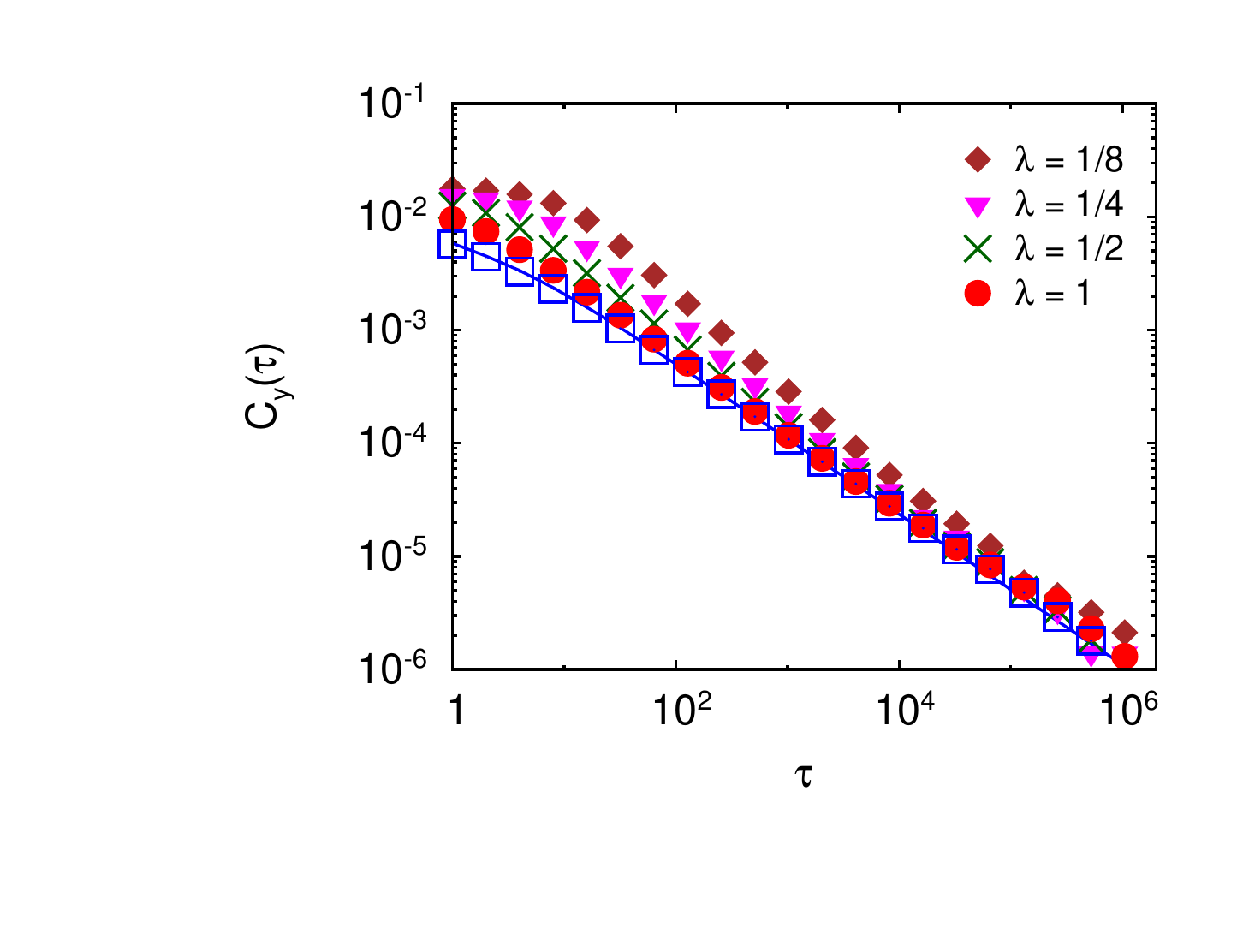} 
	\caption{%
		The figure shows the autocorrelation function of the output
		of a Preisach model with constant Preisach density
		driven by Markovian input processes with exponentially 
		decaying autocorrelation functions. 
		The parameter determining the effective Preisach density is given by 
		$\gamma_2=1/6$ and the input correlation decay rate takes the values $\lambda=1,1/2,1/4$ and $1/8$.
		The output correlation decay is compared with the corresponding situation 
		with uncorrelated driving, simulation (blue boxes) and rigorous results 
		(blue line).
	}
\label{fig:cor_P2Deq_uni_plH_AR}
\end{figure}

We consider two scenarios. At first, $\gamma_2=1/6$ is fixed and the input
correlation decay rate $\lambda$ is varied.
The output autocorrelation functions $C_y(\tau)$ which follow from the simulations are shown in \figref{fig:cor_P2Deq_uni_plH_AR}. 
The results are compared with the numerical data and the rigorous result,
Eq.\,(\ref{eq:corz_mueff2D}), for the corresponding scenario with uncorrelated
driving.
The output correlation decay approaches the correlation decay resulting from
uncorrelated driving asymptotically ($\tau\rightarrow\infty$). 
Thus, the asymptotic output correlation decay is solely determined by the
effective Preisach density irrespective of the presence of input correlations.
Again, the
deviations from the corresponding results with uncorrelated driving observed for small
$\tau$-values become larger with decreasing input correlation decay rates.
%

\begin{figure}[ht]
	\centering
\includegraphics[width=\figurewidth]{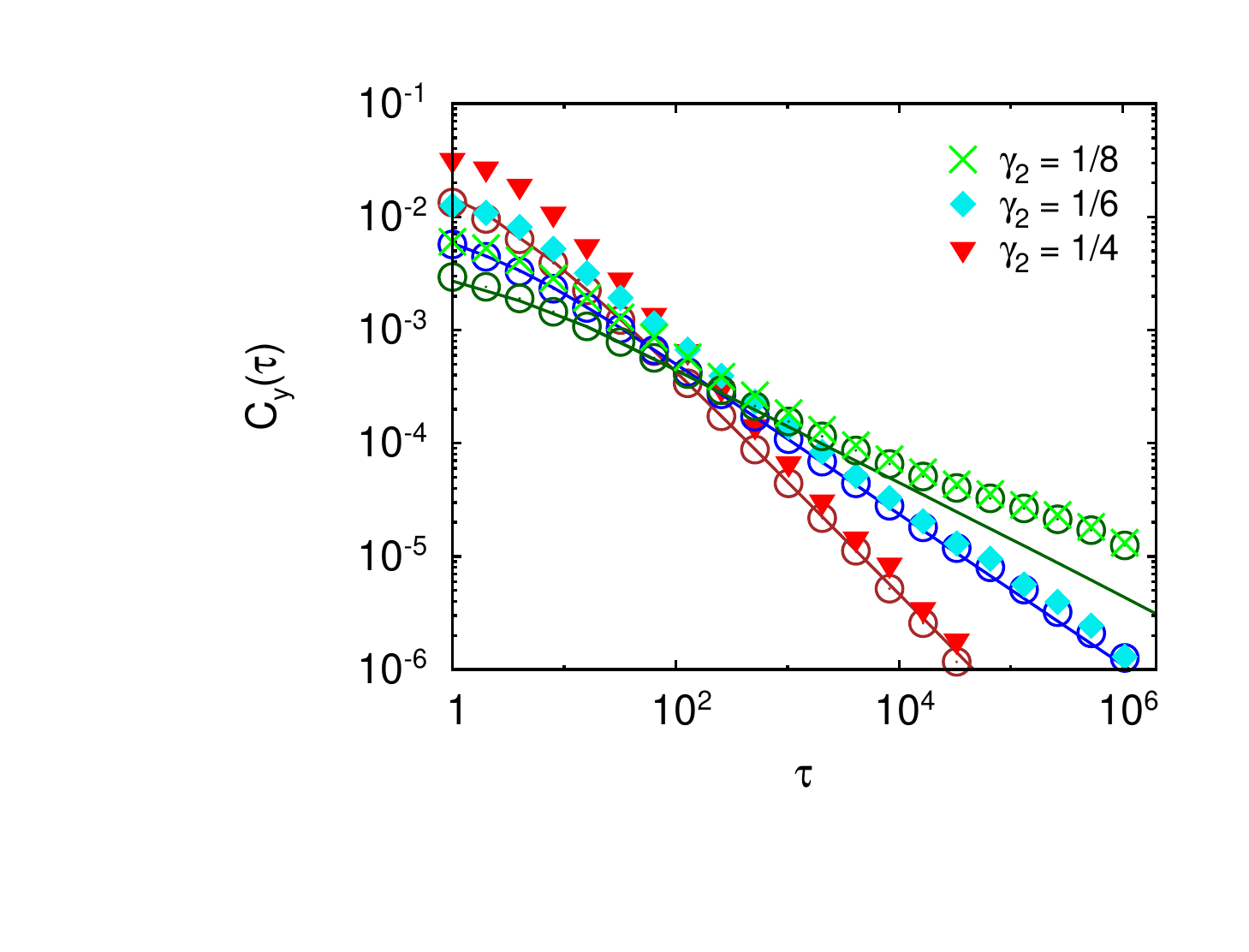}
	\caption{%
		The figure shows the output autocorrelation function 
		of a Preisach model with constant Preisach density
		driven by Markovian input processes with exponentially 
		decaying correlations. 
		The influence of input and Preisach density is captured by $\gamma_2=1/4,1/6$
		and $1/8$. The input correlation decay is determined by $\lambda=1/2$.
		The output autocorrelation functions approach the corresponding 
		correlation decay for uncorrelated driving, simulation (circles) and
		rigorous results (solid lines).
		The deviation between numerical data and rigorous result for $\gamma_2=1/8$
		is discussed in the text. 
	}
\label{fig:cor_P2Deq_uni_pl_AR02}
\end{figure}

In the second scenario, the input correlation decay rate $\lambda=1/2$ is fixed
and with $\gamma_2$ varies the broadness of the input density. The results
obtained numerically for the autocorrelation function of the hysteretic response
for three different $\gamma_2$-values are presented in
\figref{fig:cor_P2Deq_uni_pl_AR02}.
%
As observed for symmetric Preisach models, the output processes show the
asymptotic correlation decays $C_y(\tau)\sim\tau^{-\eta_\delta}$ already
known for Preisach models driven by uncorrelated input processes.  
The expected correlation decay becomes slow in case of small $\gamma_2$-values.
Consequently, numerical data of the output autocorrelation function are
more likely subject to finite-time effects.
Thus, we attribute it to finite-time effects that the values of
the autocorrelation function at large $\tau$-values are overestimated for
$\gamma_2=1/8$ in \figref{fig:cor_P2Deq_uni_pl_AR02}. 
This assumption is supported by the fact that results based on
correlated as well as uncorrelated inputs are affected in the same way. 
The numerical data of the autocorrelation function of the Preisach model with
correlated input, however, approach the corresponding data of the Preisach model
with uncorrelated input.
%
%
%

\subsection{\label{subsec:out_pdf_AR}The output density of symmetric
Preisach models}

\begin{figure}[t]
\textbf{(a)}
\includegraphics[width=\figurewidth]{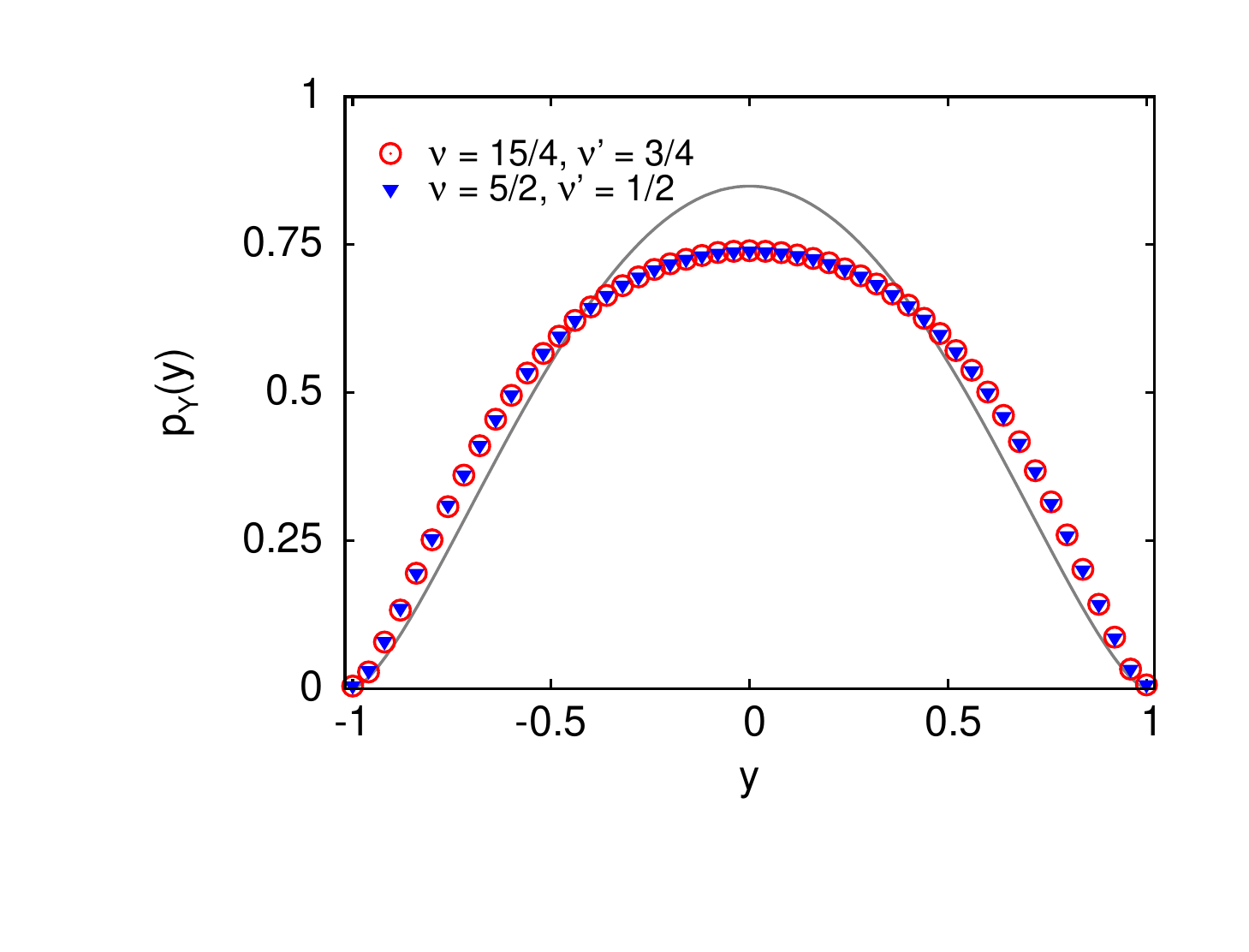}
\\
\textbf{(b)}
\includegraphics[width=\figurewidth]{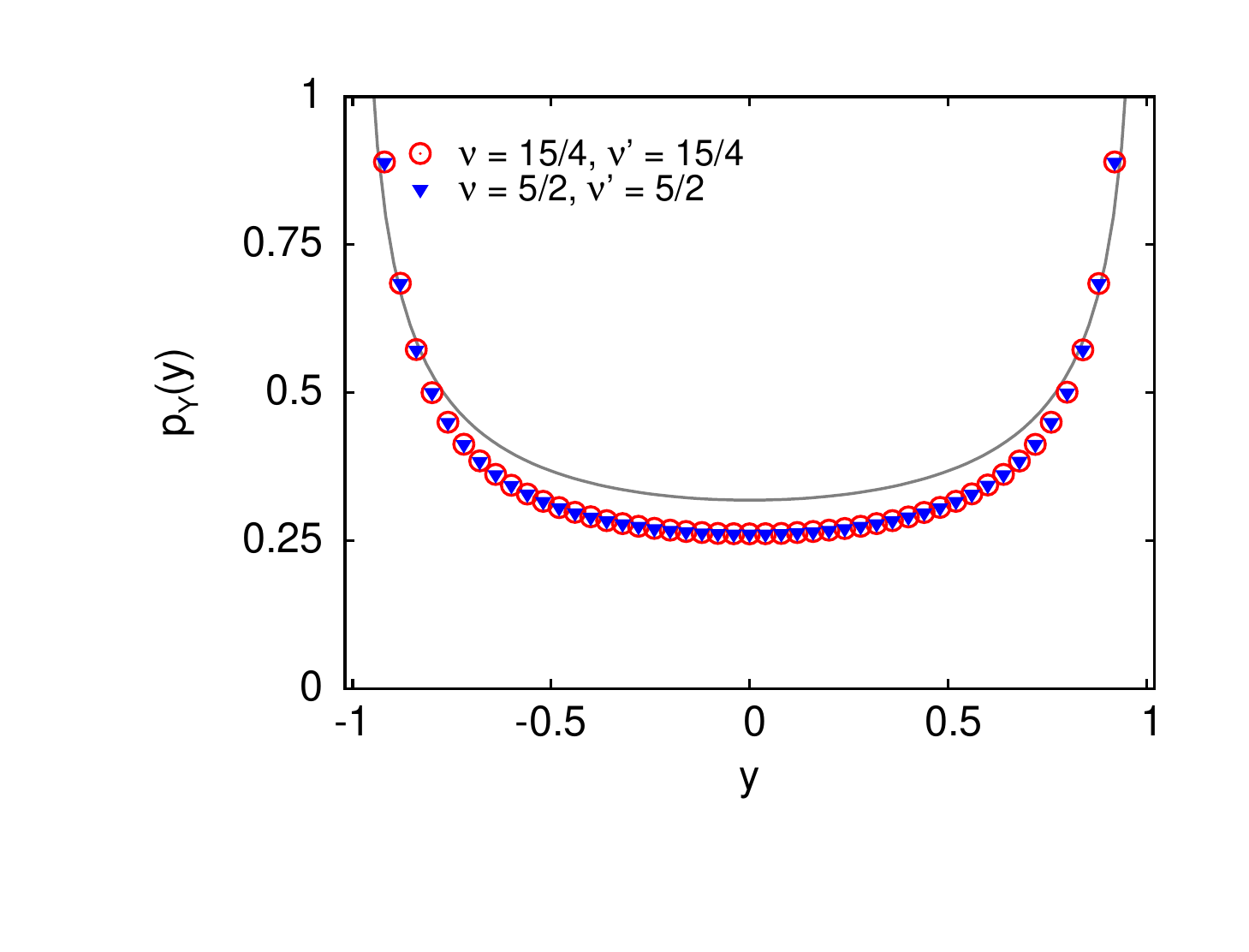}
	\caption{%
		The numerical data (symbols) of the output densities of 
		symmetric Preisach models with correlated stochastic inputs are shown.
		The input correlation decay rate takes the value $\lambda= 1/2$.
		The parameter of the effective Preisach density is given by 
		\textbf{(a)} $\gamma_1=1/5$ and \textbf{(b)} $\gamma_1=1$.
		The numerical data coincide if the models share the same effective
		Preisach density.
		The output densities, however, deviate considerably from the corresponding
		output densities of Preisach models with uncorrelated inputs (solid line), see
		Eq.\,(\ref{eq:PDFy}).
	}
	\label{fig:PDF_P1D_algB_tf02_AR} 
\end{figure}

\begin{figure}[t]
\textbf{(a)}
\includegraphics[width=\figurewidth]{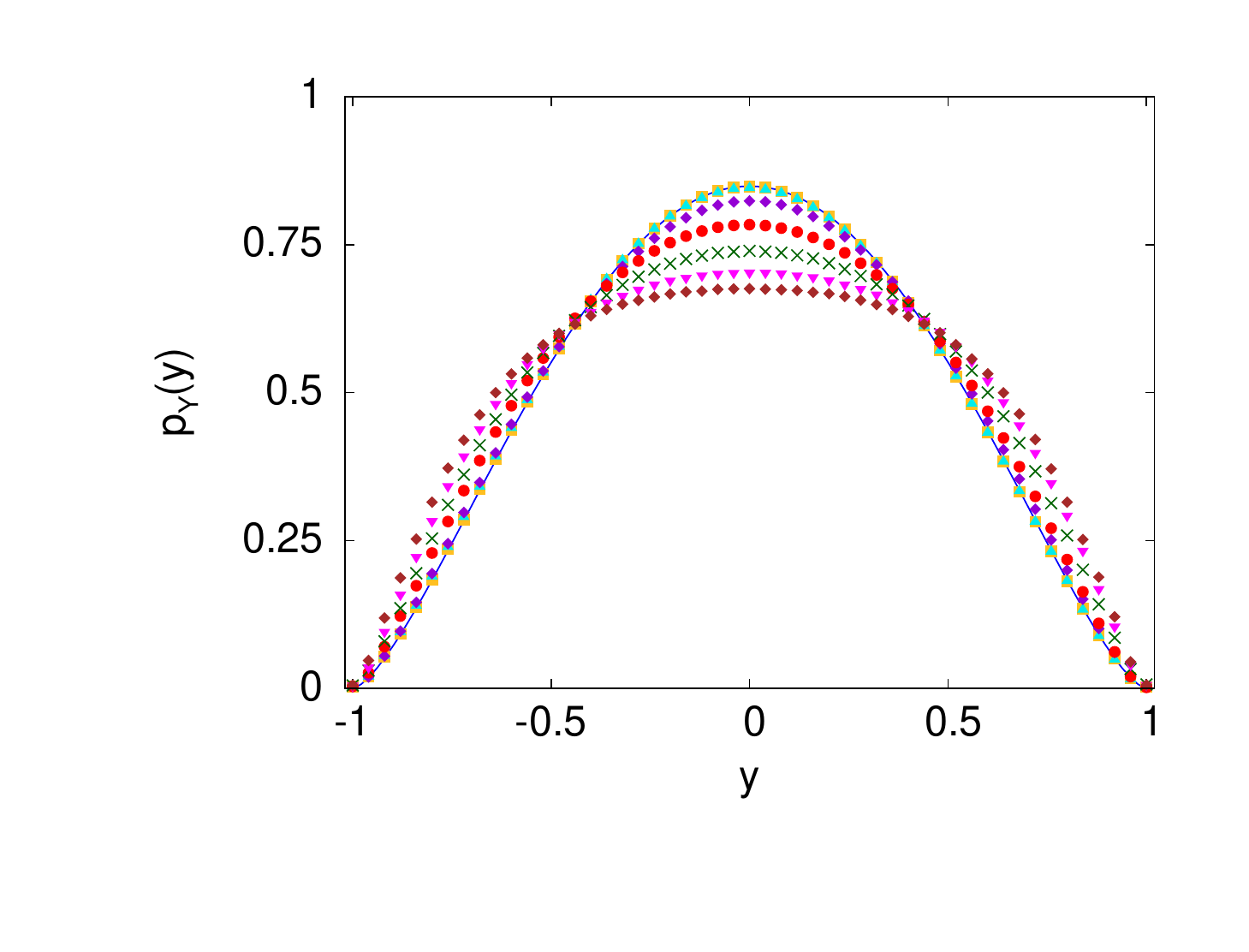}
\\
\textbf{(b)}
\includegraphics[width=\figurewidth]{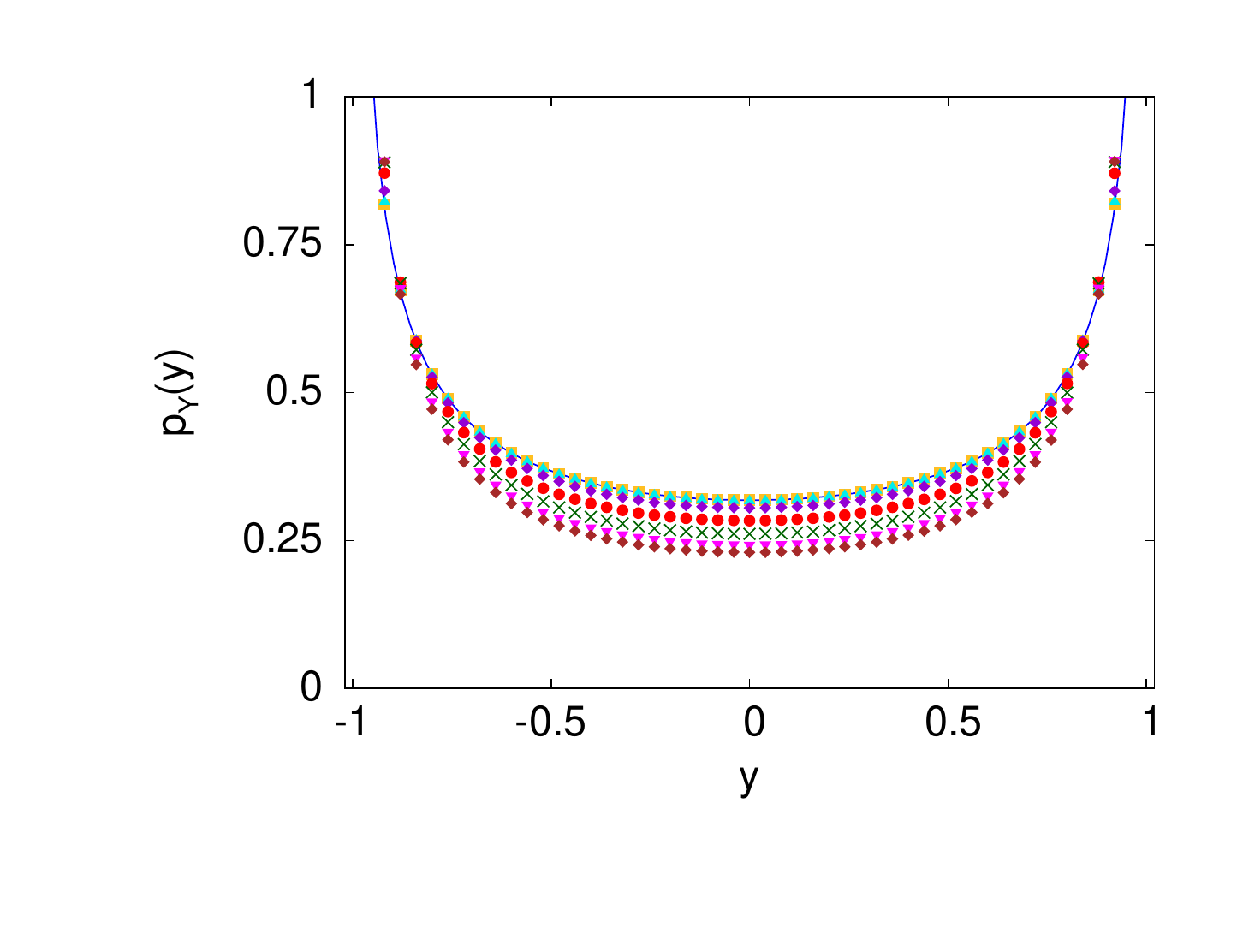}
	\caption{%
		The numerical data of the output density of 
		a symmetric Preisach model with stochastic input is shown, \textbf{(a)}
		$\gamma_1=1/5$ and \textbf{(b)} $\gamma_1=1$.
		The correlation decay rates take the values $\lambda= 1/8,1/4,1/2,\ldots,16$
		(from bottom to top near $y=0$).
	    If the input correlation decay rate becomes larger, the output density
	    becomes more narrow and closer to the output density of the model with
	    uncorrelated input (solid line).
	}\label{fig:PDF_P1D_algA_tf01_AR} 
\end{figure}

\begin{figure}[t]
\includegraphics[width=\figurewidth]{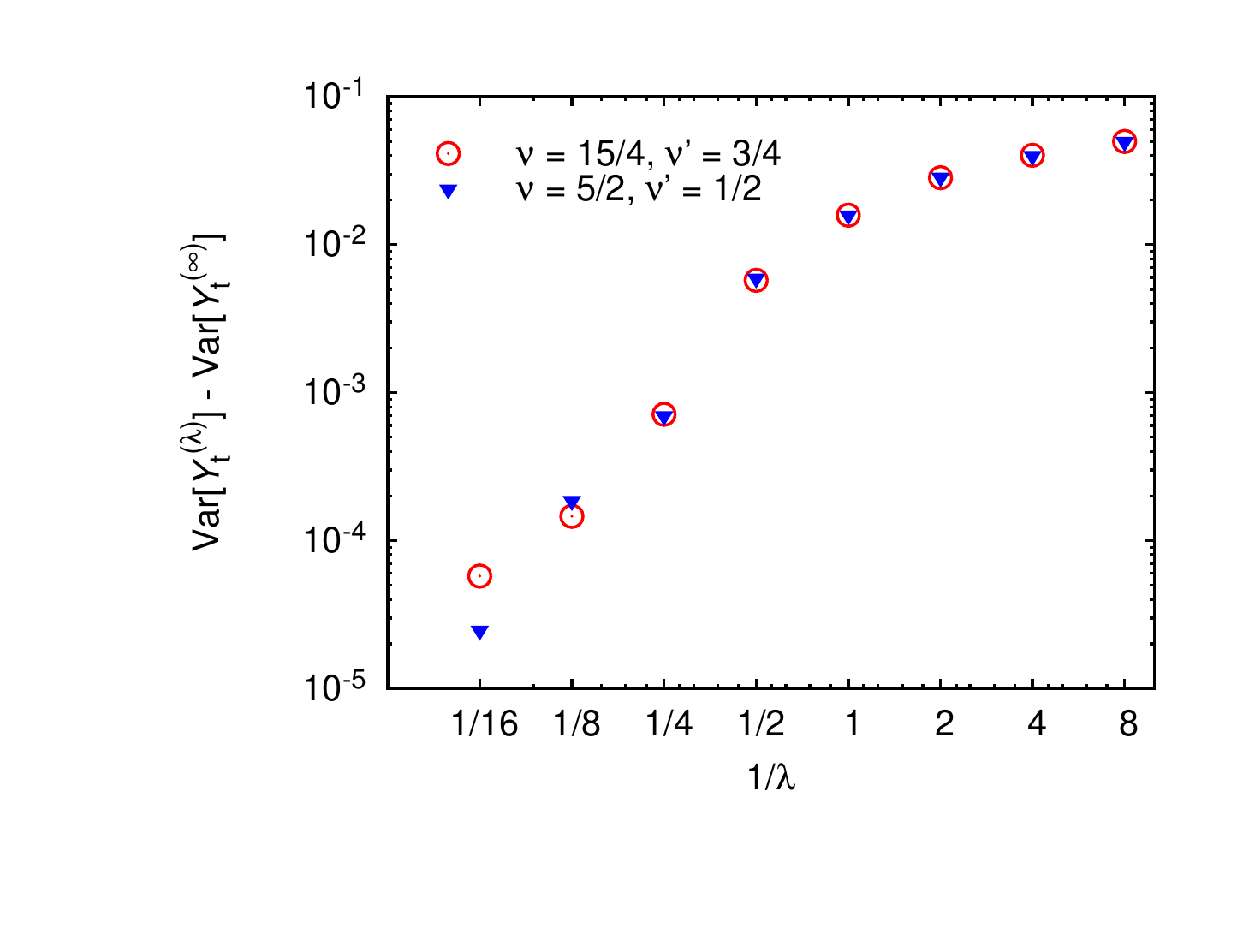}
\caption{%
		The output variance $\text{Var}[Y_t^{(\lambda)}]$ decreases with increasing
		correlation decay rate $\lambda$, here $\gamma_1=1/5$.
		It approaches the output variance of a Preisach model with uncorrelated input
		$\text{Var}[Y_t^{(\infty)}]$. 
		We attribute the small differences between the two data sets (red circles and
		blue triangles) to the finite sample size.
	}\label{fig:var_P1D_algB_tf02_AR}
\end{figure}

In this paragraph, we are going to study the effect of input correlations on the
output probability density.
We study the symmetric Preisach model with input and Preisach density according
to Eqs.\,(\ref{eq:p_algden}) and (\ref{eq:mu_algden}), respectively.
To begin with, we find
that input and Preisach density do not influence the output density
individually, but only in its combination as captured by the effective Preisach
density.
The corresponding numerical data of two scenarios
yielding the same effective Preisach density coincide, see
\figref{fig:PDF_P1D_algB_tf02_AR}.
Thus, the effective Preisach density together with the input
autocorrelation function determine the output density. 
%
To show how, we
will present some numerical results for the output density of a Preisach model
with inputs which show different correlation decays at first. Secondly, we will
give an explanation of the observed behavior. 
We look at two examples. In the first example, the parameters are $\nu=5/2$ and
$\nu'=1/2$, hence $\gamma_1=1/5$. 
In the second example, the parameter of the Preisach density is given by
$\nu'=5/2$, hence $\gamma_1=1$. 
Further, the input correlation decay rate
$\lambda$ is varied. The scenario with uncorrelated input corresponds to
$\lambda\rightarrow\infty$.
The slower the input correlation decay becomes, the more the output density
differs from the output density of the Preisach model with uncorrelated input;
the ouput density becomes broader, see \figref{fig:PDF_P1D_algA_tf01_AR}.
The latter is also illustrated by the dependence of the output variance on the
input correlation decay rate, see \figref{fig:var_P1D_algB_tf02_AR}.

Since the input density remains the same when $\lambda$ is varied, the different
output densities are a result of the different orders in which the input values appear. 
The increase of the autocorrelation function $C_X(\tau)$ is caused by a higher
conditional probability that, for instance, a large positive value $x_t$ is
followed by an other large positive value $x_{t+\tau}$. As a consequence, values
of the same sign tend to cluster more. 
We consider symmetric Preisach models
where 
the occurrence of a small positive input value
following a large positive input value 
leads to no change in the output.
As a result, the output caused by smaller input values is more often the same as
for large values.
Small input events are in the shadow of the larger events of the same sign and,
as a consequence, less noticed by the Preisach model \footnote{%
This argument is also valid for generic Preisach models for which Preisach units
with upper and lower threshold values close to each other have only little
weight.},
which yields a broader output probability density.

\section{\label{sec:sum}Summary and discussion} 
We studied the response of Preisach models to Markovian input processes.
%
At first, analytical results were presented for Preisach models driven by 
uncorrelated processes. We showed that 
the responses of systems which share an effective Preisach density yield
realizations of the same stochastic process. 
Consequently, it is sufficient to look at the effective Preisach density
introduced in ref.~\cite{Radons2008b,Radons2008c} to compute output
autocorrelation functions.
The effective Preisach density captures all relevant features from input and
Preisach density.
%
Completing our former research on spectral properties of Preisach models driven
by
uncorrelated processes, we analyzed
the generic Preisach model within the framework presented in
ref.~\cite{Radons2008b,Radons2008a}.
It was found that the visibility of long-term memory in the output
correlations is pronounced less in the generic case.
Therefore, for symmetric and generic Preisach models to show the same long-term
correlation decay, elementary hysteresis loops of large width have to show a higher weight or
extreme input events with an amplitude close to the threshold values of these
loops have to be much more rare in the generic case than in the symmetric case.  
%
Further, we showed that, 
independent of the shapes of input and Preisach density,
the autocorrelation functions and power spectral densities of the responses of
Preisach models to sequences of i.i.d.\;random variables decay monotonically. 
Also we were able to find an expression for the output density based on
empirical data for all problems covered by some standard effective Preisach
density.


Secondly, 
we considered first-order Markovian input processes 
with a finite correlation time scale 
under hysteresis represented by the Preisach model. 
This extends the research of hysteretic systems driven by uncorrelated inputs to
input processes with exponentially decaying autocorrelation functions.
%
%
Our numerical experiments indicate that 
the rigorous results for the correlation decay of the response of Preisach
models driven by uncorrelated input processes
\cite{Radons2008a,Radons2008b,Radons2008c} hold asymptotically also in the
presence of exponentially decaying input autocorrelation functions.
Solely the effective Preisach density determines the output correlation decay
asymptotically; the influence of fast decaying temporal input correlations dies
out on small time scales.
%
Thus, the mechanisms found for uncorrelated driven systems also apply in these cases and
long-time tails in the autocorrelation function of the
system response will be observed if we have broad Preisach densities assigning a
high weight to elementary loops of large width such that rare extreme events of
the input time series give a significant contribution to the output for a long period of time. 

We found numerical evidence that the output density becomes broader
with the presence of input correlations. The smaller the correlation decay rate
becomes, the broader the output density.
Furthermore, the shape of the output density does not depend on input and
Preisach density individually, but on the effective Preisach density.
Only the effective Preisach density and the input autocorrelation function
determine the output probability density -- a fact which emphazises the
important role of the effective Preisach density even for Preisach models with
correlated inputs.

\appendix

\section{\label{sec:input}Input signals -- generation and properties} 
\subsection{Gaussian processes with exponentially decaying correlations}
For our simulations, 
an exponential correlation decay $C_X(\tau)\sim e^{-\lambda\tau}$ ($\tauinfty$)
is generated using an autoregressive model of order 1 \cite{Fuller1996}. 
A realization of the AR(1)-process $\{\xi_t\}$ is given by the recursive formula
\begin{equation}\label{eq:AR1_recursive}
\xi_{t+1} = \varphi \xi_t + \sqrt{D}\varepsilon_t \text{~where~}\varphi\in[0,1)
.
\end{equation}
$\varepsilon_t\sim N(0,1)$ are 
i.i.d.\;Gaussian variables of zero mean and unit variance.

The AR(1)-process $\{\xi_t\}$ is a Gaussian process. 
For large times $t$
the process becomes stationary 
and its autocorrelation function is given by
\begin{equation}\label{eq:AR_corr}
C_\xi(\tau) = \lim\limits_{t\rightarrow\infty}
\big(\mean{\xi_t\,\xi_{t\!+\!\tau}} -\mean{\xi_t}\mean{\xi_{t+\tau}}\big) = D\frac{\varphi^\tau}{1-\varphi^2}.
\end{equation}
By choosing the parameters
\begin{equation}\label{eq:AR_para}
 \varphi=\exp(-\lambda),~ \xi_0=0 \text{~and~} D=1-\exp(-2\lambda),
\end{equation}
we determine the correlation decay rate $\lambda$ and ensure the processes to have unit variance.

\subsection{Transformation to non-Gaussian processes} 
Non-Gaussian stochastic input processes $(X_1,X_2,\ldots)$ with exponential
correlation decay and cumulative distribution function $F_X(x)$ are obtained by
monotonic transformations $X_t=f(\xi_t)$ of the AR(1)-processes $\{\xi_t\}$,
such that the cumulative distribution function $F_\xi(\xi)$ of the Gaussian
variable $\xi_t$ and $F_X(x)$ 
coincide,  
\begin{equation}\label{eq:CDF_map}
f(\xi) = F^{-1}_X [F_\xi(\xi)].
\end{equation}
Following the arguments presented in App.\,\ref{sec:trafo},
the asymptotic correlation decay of the Gaussian process $\{\xi_t\}$
is preserved under the monotonic transformation (\ref{eq:CDF_map}),
hence $C_X(\tau)\sim a_1^2 C_\xi(\tau)$ ($\tauinfty$); and 
the constant of proportionality $a_1^2$ is given by Eq.\,(\ref{eq:a_cory}).
This method thus provides a scheme for controlling probability
density and asymptotic correlation decay of a random process $(X_1,X_2,\ldots)$,
which we use for numerical experiments concerning hysteresis. 

\begin{figure}[ht]
	\centering
\includegraphics[width=\figurewidth]{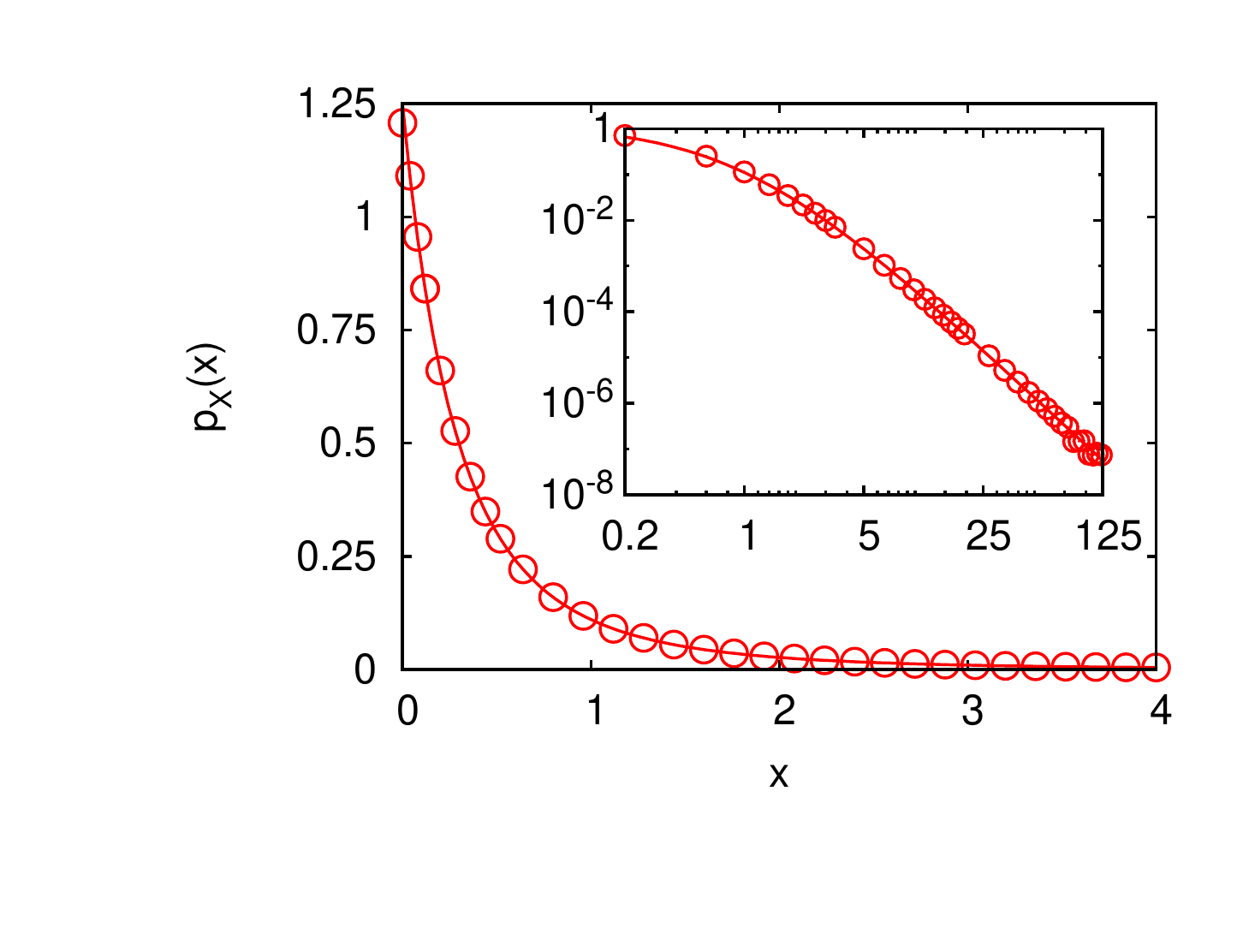}
	\caption{%
		The figure shows simulation data (circles) and analytic expression (solid
		lines) of the probability density, Eq.\,(\ref{eq:p_algden}) with $\nu = 5/2$,
		of the input process generated by the presented algorithm.
		The inset illustrates on a log-log scale
		the asymptotically algebraic decay of the 
		non-Gaussian process.
%
	}
\label{fig:wsk_tf01}
\end{figure}
\begin{figure}[ht]
	\centering
	\textbf{(a)}\hspace{-6mm}%
\includegraphics[width=\twofigurewidth]{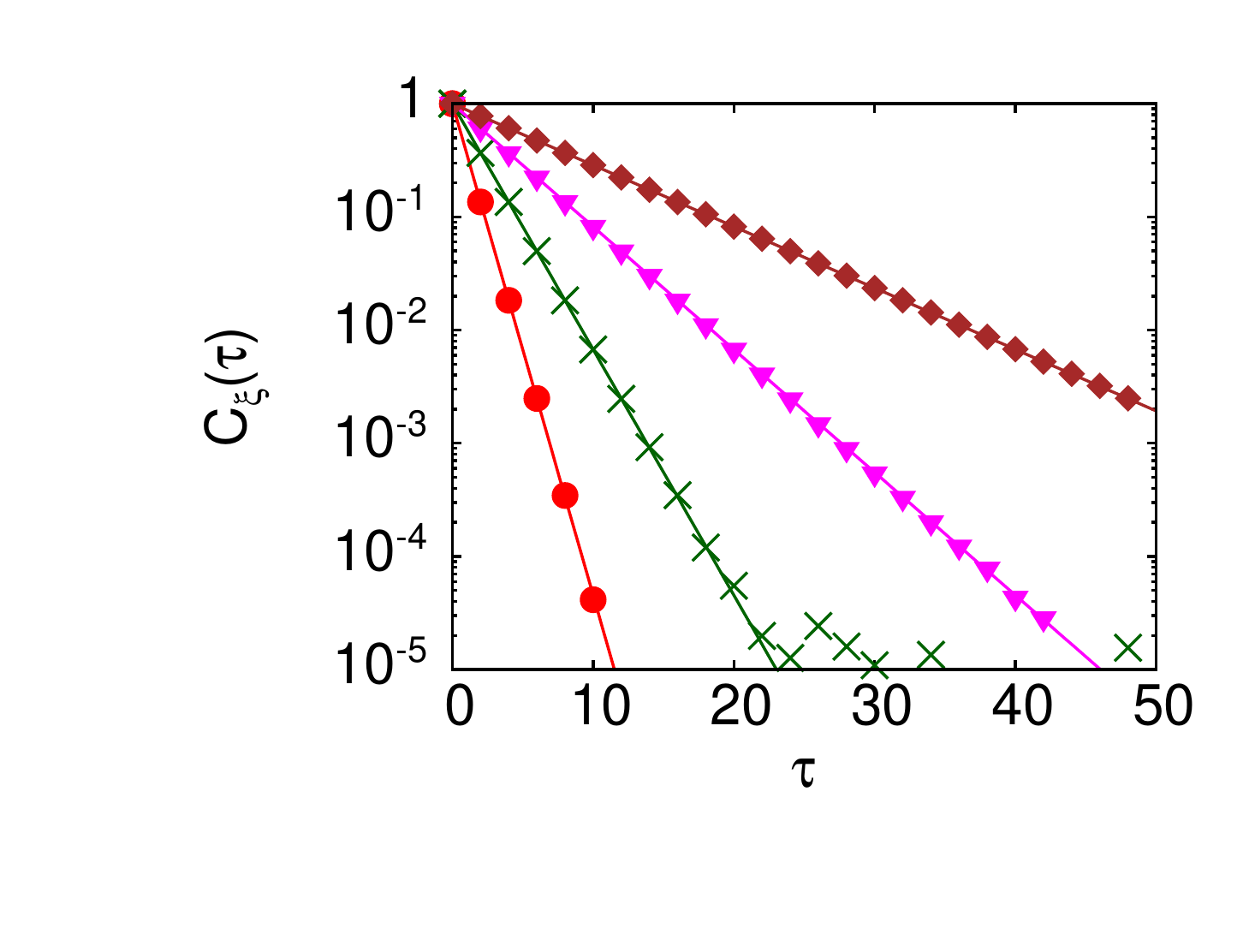} 
	\textbf{(b)}\hspace{-6mm}%
\includegraphics[width=\twofigurewidth]{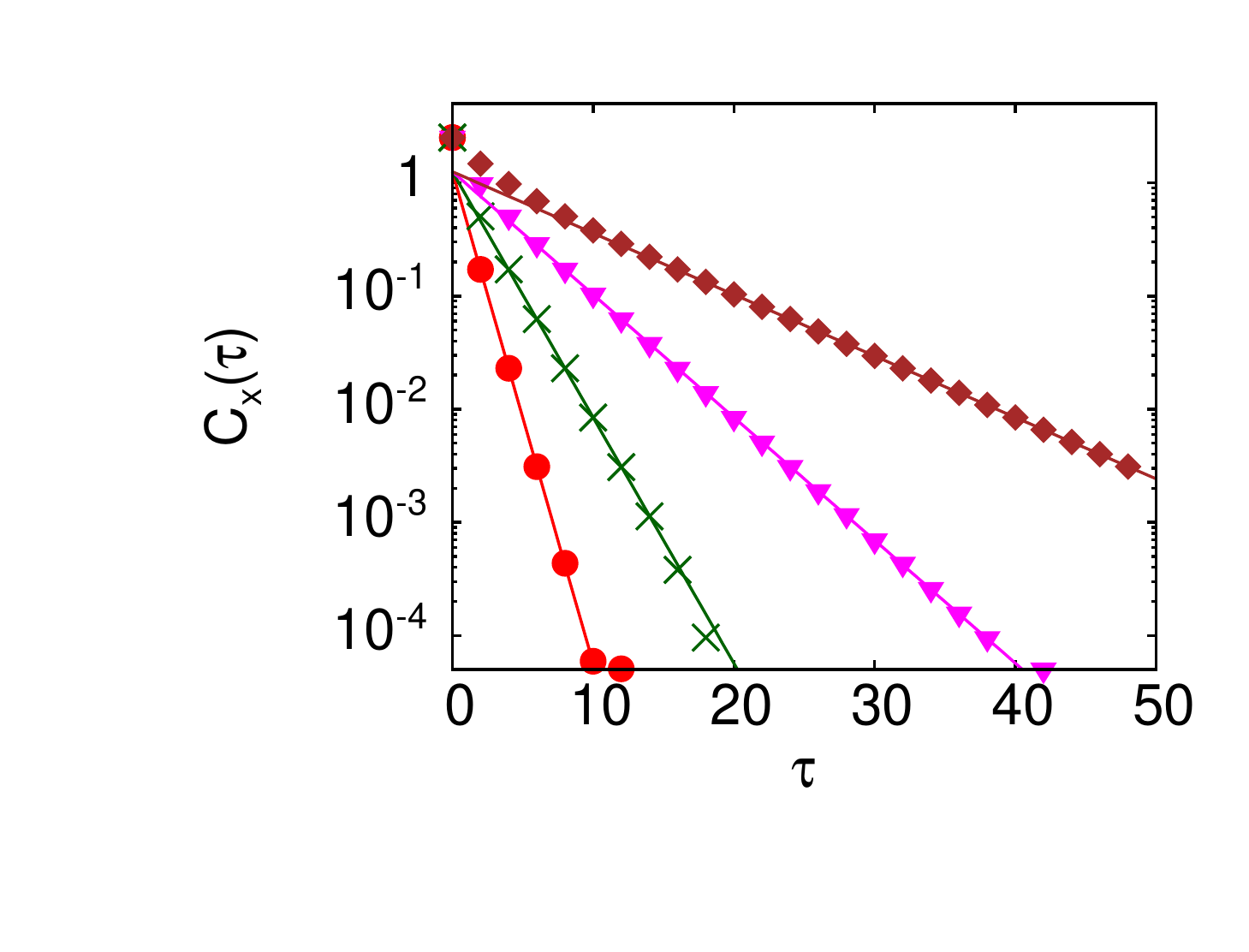}
	\caption{%
 \textbf{(a)}
		The points show numerically obtained results for the autocorrelation functions $C_\xi(\tau)$ 
		of Gaussian AR(1)-models
		with the correlation decay rates $\lambda=1,1/2,1/4$ and $1/8$. 
 \textbf{(b)}
		The non-Gaussian input processes, Eq.\,(\ref{eq:p_algden}) with $\nu = 5/2$, 
		generated by a monotonic nonlinear transformation
		show the same correlation decay. 
	}
\label{fig:cor_tf01_AR}
\end{figure}

The input probability density of one of the examples used below shows an
asymptotically algebraic decay, see Eq.\,(\ref{eq:p_algden}), 
and asymptotically exponentially decaying autocorrelation functions $C_X(\tau)\sim e^{-\lambda\tau}$.
Numerical examples are shown in Figs.\,\ref{fig:wsk_tf01} and \ref{fig:cor_tf01_AR}, respectively.
The solid lines in Fig.\,\ref{fig:cor_tf01_AR}\,b) correspond to the predicted asymptotic model behavior.

\section{\label{sec:trafo}Transformations of Gaussian processes}
Following ref.~\cite{Palma2007}, let $\{\xi_t\}$ be a Gaussian process 
with zero mean and unit variance, hence $p_\xi(\xi)=\exp(-\xi^2/2)/\sqrt{2\pi}$, and
with the autocorrelation function
$C_\xi(t,t\!+\!\tau)=\mean{\xi_t\,\xi_{t+\tau}}-\mean{\xi_t}\mean{\xi_{t+\tau}}$.
%
Further, the transformation $X_t=f(\xi_t)$ is 
a square-integrable function $\int\!\text{d}\xi\, p_\xi(\xi) |f(\xi)|^2 < \infty$
\footnote{For the variable transformation to be square-integrable, the 2nd moment of $X$ has to exist, $\mean{X^2}<\infty$.}
with Hermite expansion
\begin{equation}\label{eq:Hermite_exp}
 f(\xi)=\sum\limits_{n=0}^{\infty}\frac{a_n}{n!}H\!e_n(\xi),\,
a_n=\int\! \text{d}\xi\,p_\xi(\xi)H\!e_n(\xi)f(\xi)
\end{equation}
where $H\!e_n(\xi)$ denotes the Hermite polynomial of order $n$
\[
H\!e_n(\xi)=(-1)^n e^{\xi^2/2}\frac{\text{d}^n}{\text{d}\xi^n}e^{-\xi^2/2}.
\]
Hermite polynomials are 
orthogonal
$\int\!\text{d}\,\xi p_\xi(\xi) H\!e_n(\xi)H_m(\xi)  = n!\delta_{nm}$.
Thus, the joint Gaussian distribution $p(\xi_1,t;\xi_2,t\!+\!\tau)$ may be written as
\[
p(\xi_1,t;\xi_2,t\!+\!\tau)=
p_\xi(\xi_1)p_\xi(\xi_2)
\sum\limits_{n=0}^{\infty}\frac{C_\xi^n(t,t\!+\!\tau)}{n!}H\!e_n(\xi_1)H\!e_n(\xi_2).
\]
This becomes obvious by calculating the marginal distributions
$p(\xi_{1/2}) = \int\!\text{d}\xi_{2/1} p(\xi_1,t;\xi_2,t\!+\!\tau)$
and the autocorrelation function
$
C_\xi(t,t\!+\!\tau) = \iint\!\text{d}\xi_1 \text{d}\xi_2 \xi_1\xi_2\, p(\xi_1,t;\xi_2,t\!+\!\tau). 
$
Consequently, the autocorrelation function of the transformed process $\{X_t\}$ follows 
\begin{eqnarray*}
C_X(t,t\!+\!\tau) & = &
\iint\! f(\xi_1)f(\xi_2) p(\xi_1,t;\xi_2,t\!+\!\tau) \text{d}\xi_1 \text{d}\xi_2 -\mean{f(\xi)}^2
\\
& = &
\sum\limits_{n=1}^{\infty}\frac{a_n^2}{n!}C_\xi^n(t,t\!+\!\tau).
\end{eqnarray*}

Thus, for monotonically decreasing functions $|C_\xi(t,t\!+\!\tau)|$ with $\lim_{\tauinfty}C_\xi(t,t\!+\!\tau)=0$, the autocorrelation $C_X(t,t\!+\!\tau)$ is asymptotically $(\tauinfty)$ dominated by the smallest power of $C_\xi(t,t\!+\!\tau)$.
Consequently, we can truncate the series after the linear term
\begin{equation}\label{eq:cor_y_int_a}
C_X(t,t\!+\!\tau) \sim a_1^2 C_\xi(t,t\!+\!\tau)
\;(\tauinfty)
\end{equation}
with $a_1 = \int\!\text{d}\xi\,p_\xi(\xi)f(\xi)H\!e_1(\xi)$ where $H\!e_1(\xi)=\xi$.
As a consequence, the asymptotic correlation decay of a Gaussian process $\{\xi_t\}$
is preserved under the transformation $X_t=f(\xi_t)$ 
if $a_1 \neq 0$.
In case $\lim_{\xi\rightarrow\pm\infty}f(\xi)p_\xi(\xi)=0$,
using integration by parts,
it follows 
\begin{equation}\label{eq:a_cory}
a_1=\int\!\text{d}\xi\,p_\xi(\xi)f'\!(\xi).
\end{equation}
Thus, one can easily see that $a_1\neq 0$ is always valid
for a monotonic transformation for which $f'\!(\xi)\geq 0$ $\forall \xi$.
{The trivial case $f(\xi)=$\,const. is excluded from this statement}.

\section{\label{subsec:eff_den}Role of the effective Preisach density}
The effective Preisach 
density defined in Eq.\,(\ref{eq:def_mueff}) for the symmetric Preisach model
and in Eq.\,(\ref{eq:def_mueff2D}) in general 
provides more than expressions for the autocorrelation functions and power spectral densities of the system response to uncorrelated inputs.
Considering model systems consisting of a sequence of i.i.d.\;continuous random variables $\{X_n\}$ (uncorrelated input process) with probability density $p_X(x)$ and a Preisach model with Preisach density $\mu(\alpha,\beta)$ and a certain initial configuration of the Preisach units, we can show:
All model systems which share the same effective Preisach density $\tilde\mu(u,v)$ and are, for simplicity, initialized with the equilibrated initial state (see Sect.\,\ref{sec:model})
generate the same output sequence of random variables $(Y_1,\ldots,Y_t)$.


We are going to clarify this fact 
for the symmetric Preisach model only. One can follow the same outline for the
generic Preisach model.
We consider two model systems sharing the same effective Preisach density $\tilde{\mu}(u)$ with input and Preisach densities $p_1(x)$, $\mu_1(\alpha)$ and $p_2(x)$, $\mu_2(\alpha)$, respectively. 
There exists an equally probable realization $(x_1^{(1)},\ldots,x_t^{(1)})$ of the input process $\{X_n^{(1)}\}$
for each realization $(x_1^{(2)},\ldots,x_t^{(2)})$ of the input process $\{X_n^{(2)}\}$, 
such that the output trajectories $(y_1,\ldots,y_t)$ of both systems coincide.
%

%

The probability of a certain input trajectory is given by $p_1(x_1^{(1)})\ldots p_1(x_t^{(1)})\,\text{d}x_1^{(1)}\ldots \text{d}x_t^{(1)}$ and $p_2(x_1^{(2)})\ldots p_2(x_t^{(2)})\,\text{d}x_1^{(2)}\ldots \text
{d}x_t^{(2)}$, respectively.
An equally probable realization of the process $\{X_t^{(1)}\}$ follows from a monotonic transformation of the realization of the process $\{X_t^{(2)}\}$ so that their cumulative distribution functions coincide $F_1(x_n^{(1)})=F_2(x_n^{(2)})$,
\begin{equation}\label{peq:CDF_map}
x_n^{(1)}=f(x_n^{(2)}) \text{~where~} f(x)=F_1^{-1}[F_2(x)].
\end{equation}
As a consequence, the input densities transform as 
\begin{equation}\label{peq:PDF_trafo}
p_2(x)=p_1[f(x)]f'(x).
\end{equation}
At first, we claim that the effective Preisach densities are the same
\[
 \frac{\mu_2[\alpha_2(u)]}{2\,p_2[\alpha_2(u)]}\stackrel{!}{=}
 \frac{\mu_1[\alpha_1(u)]}{2\,p_1[\alpha_1(u)]} .
\]
Using Eq.\,(\ref{peq:PDF_trafo}), one can write
\begin{equation}\label{peq:step1}
 \mu_2[\alpha_2(u)] =
 \frac{p_1\{f[\alpha_2(u)]\}f'[\alpha_2(u)]}{p_1[\alpha_1(u)]}
 \mu_1[\alpha_1(u)] .
\end{equation}
With $u = 2[1-F_i(\alpha_i)]$ for $i=1,2$ follows $F_1(\alpha_1)=F_2(\alpha_2)$,
hence
\begin{equation}\label{peq:alpha_map}
\alpha_1=f(\alpha_2). 
\end{equation}
The latter plugged into Eq.\,(\ref{peq:step1}) yields
\begin{equation}\label{peq:step2}
 \mu_2(\alpha_2) = \mu_1[f(\alpha_2)]f'(\alpha_2) .
\end{equation}
Next, we take a look at 
the response of the Preisach model 
which is given by a sum of certain integrals, see Sect.~\ref{subsubsec:out_sim}.
Using Eqs.\,(\ref{peq:CDF_map}),(\ref{peq:alpha_map}), and (\ref{peq:step2}), one can easily show 
\[
\int\limits_{0}^{x^{(2)}}\!\text{d}\alpha_2\,\mu_2(\alpha_2) = 
\int\limits_{0}^{x^{(1)}=f(x^{(2)})}\!\text{d}\alpha_1\,\mu_1(\alpha_1) .
\]
Thus, corresponding values of both input trajectories contribute with the same value to their system's response. Consequently, both trajectories yield the same output realizations $(y_1,\ldots,y_t)$, 
assuming the Preisach units are initialized appropriately.
Model systems with the same effective Preisach density $\tilde\mu(u)$ yield the same random output process $\{Y_n\}$,
and, of course, the same autocorrelation function $C_Y(\tau)$.

As a result of this, systems with
broader Preisach densities assigning a higher weight to Preisach units of larger widths are equivalent to systems with more narrow input densities causing extreme events to be more rare.
Consequently, one needs to discuss effective Preisach densities only.

\newpage 
\bibliography{hysteresis,stochastic}

\end{document}